\documentclass[a4paper,11pt]{article}
\pdfoutput=1
\usepackage{mymacros}

\title{Islands and entanglement entropy in $d$-dimensional curved backgrounds}
\author{Filip Landgren, Arvind Shekar}
\affiliation{Mathematical Sciences and STAG Research Centre, University of Southampton,\newline Highfield, Southampton, SO17 1BJ, UK.
}

\emailAdd{f.landgren@soton.ac.uk, a.shekar@soton.ac.uk}

\begin{document}

\abstract{A large part of the discussion on entanglement islands has explored the specific setup of $2d$ JT gravity with a flat heatbath coupled to a $2d$ CFT.
In this paper, we consider a more general setup and treatment of islands in a $d-$dimensional AdS black hole background. The quantum fields modeling the Hawking radiation have a scale and are consistently inherited from a conformal parent theory; their symmetries are thus compatible with those of curved backgrounds. We demonstrate explicitly that the existence of islands is sensitive to the choice of CFT used to model the Hawking radiation. 
We compute the renormalised entanglement entropy of conformal fields on a negatively curved background in $d$ dimensions at zero temperature as well as the thermal regulated entropy of an entangling region near the UV boundary. Using the latter quantity as the entropy of the Hawking radiation, we find that islands never emerge for $d>2$.}

\maketitle
\newpage
\section{Introduction}
Entanglement entropy has been well studied in two dimensions but is still much less understood in higher dimensions; see, for instance, the works of \cite{Casini:2022rlv, Calabrese:2004eu}. Entanglement entropy is typically studied by quantum field theory methods for vacuum states of quantum field theories at zero temperature. Entanglement is of course subtle at finite temperature since quantum and thermal effects mix \cite{Cardy:2014jwa, Herzog:2014fra,Herzog:2015cxa}. Computation of entanglement entropy for generic excited states using standard quantum field theoretic methods such as heat kernels is often intractable. For free massive quantum fields one can explore entanglement entropy in curved backgrounds using heat kernel techniques \cite{Vassilevich:2003xt} but such methods are not applicable in the massless limit. 

The goal of this paper is to explore entanglement entropy in specific types of curved backgrounds, namely those related to black holes. Our motivations derive from the recent discussions of islands in the context of the black hole information loss paradox. The key idea of the islands proposal \cite{penington2020entanglement, almheiri2019islands} is the following: it is not sufficient to use the classical Bekenstein-Hawking formula to compute the entropy of a black hole spacetime, but one instead needs to use the generalized entropy formula. There are hence two distinct contributions that need to be taken into account to compute the entropy of the black hole $S_{BH}$: 
\begin{equation}
S_{BH} = {\rm Ext}_Q \left [ \frac{ { \cal{A}}(Q)}{4 G} + S_{\rm matter}(B) \right ] \label{islands1}.
\end{equation}
Here ${\cal A}(Q)$ denotes the area of a quantum extremal surface and $B$ denotes the region between the quantum extremal surface and the spacetime boundary. $S_{\rm matter} (B)$ denotes the entanglement entropy of quantum fields in the region $B$. The generalized entropy is obtained by extremising the surface $Q$ such that the entropy is minimised. 

Many of the papers discussing quantum extremal islands have explored the specific setup of two-dimensional JT gravity with the quantum fields being described by a two-dimensional conformal field theory. The use of a two-dimensional conformal field theory leads to a number of conceptual and technical simplifications, which we summarize in table \ref{tab:difference table}. Technically, the required computations of the quantum field theory entanglement entropies make use of well-studied and well-known results for two dimensional CFTs, for single and multiple intervals. Entanglement entropy for a conformal theory in two dimensions also has a number of conceptual simplifications. Firstly, a two-dimensional metric can always be parameterised in terms of a single conformal factor; this means that we can characterise entanglement in a background with a generic metric in terms of the conformal factor. Secondly, in two dimensions a black hole horizon is of dimension zero and one can use coordinate transformations to relate spacetimes with and without horizons. Accordingly, one can obtain the finite temperature entanglement from the zero temperature results using such coordinate transformations. 

Another important conceptual and computational aspect of the two dimensional computations relates to the regularisation of the quantum field theory contribution. In the condensed matter literature it is standard to work with regulated entanglement entropy, rather than renormalise; the UV regulator is related physically to the lattice scale of the system of interest. Much of the literature computing entanglement entropy therefore focuses on regulated expressions. In a two-dimensional conformal field theory the regulated divergences are logarithmic, related to the conformal anomaly; the coefficients are proportional to the central charge and these contributions are therefore often termed universal. 

The expression \eqref{islands1} is however a semi-classical gravity expression and in semi-classical gravity one usually works with renormalised quantum field theory contributions of regulated expressions. It was noted in \cite{Casini:2004bw} that finite contributions to the entanglement entropy can be obtained through differentiation with respect to a scale of the entangling region (see also \cite{ryu2006aspects} for a review). For example, for a region of scale $l$ in a two-dimensional conformal field theory, the expression 
\begin{equation}
l \frac{\partial S_{\rm CFT}}{\partial l}  \label{liu}
\end{equation}
is automatically finite because the (local) regulated UV divergences are independent of the (non-local) entangling region scale. There are analogous expressions for higher-dimensional CFTs involving higher numbers of derivatives. Separately, systematic renormalisation of quantum field entanglement entropy has been developed in \cite{Taylor:2016aoi}; in this approach the renormalisation of the entanglement entropy is inherited from the renormalisation scheme of the partition function.  

When one applies \eqref{islands1} in the context of JT gravity and two-dimensional conformal field theory, one determines the quantum extremal surface by extremising with respect to the scale of the region $B$. This extremisation is independent of the regulator because of \eqref{liu} (although the value of the actual extremised entropy in \eqref{islands1} does still depend on the regulator). This property is however specific to two-dimensional CFT: the first derivative of the entanglement entropy in a conformal field theory in $d > 2$ is in general not finite.  

\bigskip

Models of islands in higher dimensions were first discussed \cite{Almheiri:2019psy} and later in \cite{Chen:2020uac, Chen:2020hmv,Geng:2020qvw}.  These are known as double holographic models (see also \cite{Neuenfeld:2021wbl}) where gravity in an AdS$_{d+1}$ region bounded by a pair of end-of-the-world branes is dual to a CFT$_{d−1}$ theory living on the interception between the branes. These are examples of  Karch-Randall braneworld models \cite{Randall:1999vf, Randall:1999ee,Karch:2000ct, Giddings:2000mu,Karch:2000gx}. Despite the renewed interest in these models and their recent success, there are persistent challenges that arise when doing holography with AdS with a cutoff. In this paper, we will not explore these models further.
   
Explicit realization of islands in higher dimensions, without resorting to braneworld models, stems in part from the lack of analytical control over entanglement entropy of the quantum fields in $d>2$.   

In our construction, we will perform a circular uplift of NAdS$_2$ à la Kaluza Klein, which results in AdS$_3$ with a compact direction. The two-dimensional dilaton becomes a dimensionful parameter in the AdS$_3$ metric. We then place the AdS$_3$ spacetime on the boundary of AdS$_4$. Using the Ryu-Takayanagi (RT) formula \cite{Ryu:2006bv} entanglement entropy can be computed holographically by computing the area of a co-dimension two surface homologous to the entangling region. Since the fields inherited from the parent action in this model have a scale, they respect the symmetries of the background. In the setup where JT gravity with a bath is coupled to a $2d$ CFT, the explicit and spontaneous breaking of the symmetries in JT gravity renders the background incompatible with the symmetry of the $2d$ conformal fields. 

We consider an annulus entangling region that naturally captures the circular extra dimension and due to the isometry in the circular direction, the entanglement entropy will only depend on one coordinate (the width of the annulus). 

We generalize this construction to $d$-dimesnions and explicitly evaluate the renormalised entanglement entropy for zero temperature states in a black hole background. This quantity has interesting applications on its own to condensed matter physics and cold atom systems in particular (see \cite{Laflorencie:2015eck} for a review). In the context of islands, we are interested in the thermal entanglement entropy\footnote{As Hawking originally pointed out \cite{Hawking_2005}, an observer would need to collect measurements over an infinite time scale at spatial infinity to differentiate between a mixed or pure initial state of the black hole. In the quasi-stationary regime (on timescales small in comparison with the evaporation scale, or when the entangling region is far away from the horizon) the Hawking radiation is described by an out-of-equilibrium process and we may work with the near thermal state approximation.}. Near the UV boundary, the regulated entanglement entropy is dominated by the divergent piece whose structure is the same in the zero and finite temperature case, so we may use the covariant counterterm of the zero temperature renormalised entropy as the thermal regulated entropy. We place the entangling region near the UV boundary and demonstrate the non-generic existence of islands in higher dimensions.

\subsection{Organization of paper and summary of results}
In section \ref{sec: uplift} we perform a consistent uplift of NAdS$_2$ to pure AdS$_3$ for a non-conformal $D_p$
brane background. 

In section \ref{dimanalysis} we use dimensional analysis and symmetry arguments to constrain the form of the renormalised entanglement entropy at zero and finite temperature for an annulus entangling region. 
On a flat background, we have translational invariance and the entropy takes the form $\Sren \sim c \frac{L_y}{\Delta x}$, where $c$ corresponds to the central charge and $L_y$ is the circumference of the compact angular direction. On an AdS$_3$ background, the entropy takes the form $
\Sren \sim c \frac{L_y}{x_{1,2}} g\left( \frac{\ell_3}{x_2}, \frac{x_1}{x_2} \right)$
where $g\left( \frac{\ell_3}{x_2}, \frac{x_1}{x_2} \right)$ is some function of dimensionless ratios. 

In section \ref{sec: flat background} we compute the renormalised entanglement entropy on a flat background in three dimensions to 
\begin{equation}\label{eq: intro Sren}
S_{\textnormal{ren}}= - \frac{\pi ^2 \Gamma \left(\frac{3}{4}\right)^2}{G_4 \Gamma \left(\frac{1}{4}\right)^2} \frac{2 \phi_c \ell^2}{L}
\end{equation}
where $L$ is the width of the annulus, $\ell$ is the curvature length of the ambient space and $\phi_c$ is the periodicity of the compact direction. 

We see in section \ref{sec: EE curved background} that the entanglement entropy area functional for a curved background in Poincaré coordinates has no explicit dependence on the three-dimensional curvature length, $\ell_3$. In the flat limit, where $\ell_3 \to \infty$, we should recover the flat limit. But since $\ell_3$ is only an implicit parameter inside of the Poincaré coordinates, the entanglement entropy on a curved background in these coordinates must have the same functional form as the entanglement entropy on a flat background. Thus, in the zero temperature case, one can effectively take the flat result and swap the flat coordinates to Poincaré coordinates. In this way, we recover the result from the flat limit without having to compute the minimal area of the RT surface. This can also be seen from the perspective that all the entanglement structure is localized to a point of the $3d$ conformal boundary. In appendix \ref{sec: analysis of RT surface} we analyze the behavior of the implicit solution to the minimal surface and determine the counterterms from asymptotic analysis as a consistency check. The latter agrees with the covariant counterterms computed in  section \ref{sec: explicit counterterms} which generally takes the form
\begin{equation}
    \Sct = \frac{1}{4G_{d+1}} \int_{\d A} d^{d-1}x^\alpha \sqrt{\tilde{h}}
\end{equation}
where $\tilde{h}$ is the induced metric on the boundary of the entangling region. 

The results are generalized to $d-$dimensions in section \ref{sec: higher d}. We take the background geometry to be a $d$-dimensional AdS black hole background. Using a similar analysis as in the AdS$_3$ case, we compute the renormalised entanglement entropy on a curved background to
\begin{equation}
    \Sren = \frac{\ell_{d+1} \phi_c^{d-2}\Omega_{d-2}}{4G_{d+1}} \rho_0^{1-\frac{d}{2}} \frac{\Gamma\left[ \frac{2-d}{d(d-1)}\right] \Gamma \left[ \frac{1}{2}\right]}{\Gamma \left[ \frac{1}{d(d-1)}\right]}
\end{equation}
where $\rho_0$ specifies the turning point of the minimal RT surface and the divergence is manifestly removed with the covariant counterterm
\begin{equation}
    \Sct= -\frac{\Omega_{d-2}}{4 G_{d+1}}\left(\frac{\ell_4  \phi_c}{ \sqrt{\epsilon}} \right)^{d-2} \left( \frac{1}{x_2^{d-2}} - \frac{1}{x_1^{d-2}} \right).
\end{equation}

In section \ref{sec: islands UV} we conclude by considering islands in $d$ dimensions by placing the edge of the entangling region close to the UV boundary where the regulated thermal entanglement entropy is dominated by the divergent piece.

The regulator of the regulated entropy remembers that we are working with an effective field theory and could characterize the scale beyond which heavy modes emitted by the evaporating black hole become important.
However, we find that islands are always absent for $d>2$.

\begin{table}[H]
\begin{center}
\begin{tabular}{|| p{8cm}|p{8cm} ||} 
 \hline
\textbf{Islands in the $2d$ JT gravity setting} & \textbf{Islands in AdS$_{d \geq 3}$} \\ [1ex] 
 \hline\hline
  The matter fields are added to an already reduced gravitational theory. It is assumed that they do not couple to the dilaton: 
  \newline 
  $\mathcal{I}_{\rm JT}[\Phi, g_{\mu \nu}]+ \mathcal{I}_{\rm CFT_2}[g_{\mu \nu}]$. & The matter fields are inherited from the pure AdS$_{d\geq 3}$ action and do couple to the dilaton:
  \newline 
  $\mathcal{I}[g_{\mu \nu}, \Phi]$.  \\ 
 \hline
The matter fields are scale-invariant and do not respect the symmetries of the background. & The matter fields has generalized conformal structure and do respect the symmetries of the background. \\
 \hline
The structure of the divergence obscures if entropy is renormalised or regulated as it coincides with the logarithmic Weyl anomaly term.  & Clear distinction between regulated and renormalised quantities.  For $d>2$ the divergence follows a power law: $S \sim \epsilon_{UV}^{d−2} A_{d−2}$ where $A_{d−2}$ is the area of the boundary of the entangling region.  \\
\hline
 Black holes with horizons can be mapped to black holes without horizons. Entropy at finite and zero temperature is thus related via a coordinate transformation.  & The background has an inherent temperature in quasi-thermal equilibrium with the temperature of the thermal fields, over small timescales relative to the evaporation process.  \\
 \hline
 Heat baths collecting the Hawking radiation are required, in which an entangling region can be consistently defined due to the enhanced diffeomorphism invariance and topological nature of $2d$ gravity. & An entangling region can be consistently defined in the AdS region. We impose transparent boundary conditions at the conformal boundary but remain agnostic about what is on the other side. As elaborated on in \cite{landgren1}, all information should be retrievable within the AdS region.  \\
 \hline
The first derivative of the entanglement entropy with respect to the scale of the entangling region, $l$, say: $l \frac{\partial S_{\rm CFT}}{\partial l} $ is always finite as the regulated UV divergences are independent of the (non-local) entangling region scale.   & The first derivative of the entanglement entropy with respect to the scale of the entangling region is in general not finite. \\
\hline
Parent theory is a $4d$ near extremal black hole, with an AdS$_2 \times S^2$ near horizon region.  & Parent theory is pure AdS. \\ [1ex] 
 \hline\hline
\end{tabular}
\end{center}
 \caption{Summary of conceptual and computational differences between working with a $2d$ CFT placed on JT gravity and circularly uplifted AdS$_{d \geq3}$.}
\label{tab:difference table}
\end{table}

\section{Uplifting NAdS$_2$ to AdS$_3$ and beyond}\label{sec: uplift}
Holography for AdS$_2$ has gone through thorough investigation in recent years but remains less understood than its higher dimensional siblings. The main reason is that pure AdS$_2$ is over-constrained by its symmetries and there is no consistent notion of energy excitations above the vacuum \cite{Compere:2013bya, Castro:2014ima, Hartman:2008dq, Strominger:1998yg}. This problem was partly resolved by Almheiri and Polchinski \cite{Almheiri:2014cka} by considering the leading order correction away from pure AdS$_2$ which gives 'nearly' AdS$_2$ or just NAdS$_2$. This leading order correction away from AdS$_2$ has a universal form in the sense that the gravitational backreaction can be described by a  universal AdS$_2$ dilaton gravity. This two-dimensional dilaton gravity matches the leading non-conformal effects in the low energy limit of the SYK model \cite{Sachdev_1993, kitaev2015simple}.

The two-dimensional dilaton gravity with a Maxwell field describes the very near horizon effective theory of five dimensional nearly extremal black holes \cite{Strominger:1998yg}, and can be obtained by a consistent circle reduction from pure AdS$_3$. By letting the Maxwell field consistently vanish gives us Jackiw–Teitelboim (JT)
gravity \cite{Jackiw:1982hg, Teitelboim:1983fg} (see \cite{Mertens:2022irh} for a review) which locally has AdS$_2$ geometry with a running dilaton.

An important aspect of the holography of JT gravity is the breaking of
conformal symmetry. The asymptotic symmetries of AdS$_2$ are time reparametrizations of the boundary and are also explicitly broken in JT gravity when we consider small deformation away from pure AdS$_2$\footnote{JT gravity is topological and has no propagating degrees of freedom so the gravitational backreaction is governed by its symmetries.}. 

There is an inherent relationship between NAdS$_2$ with a dilaton and pure AdS$_3$ in which the dilaton becomes a parameter in the metric. 

The realization of a holographic description of AdS$_2 \times S^2$ or AdS$_2 \times S^3$ requires a consistent Kaluza-Klein reduction of an uplifted theory over the compact manifold. 
This was shown for Dp branes asymptotically conformal to AdS$_{p+2} \times S^{8-p}$ in \cite{Kanitscheider:2008kd}. We will consider an entangling region on a non-conformal Dp-brane background,  whose 10-dimensional action  in the dual frame takes the form \cite{Boonstra:1998mp}
\begin{equation}
    \mathcal{I}_{10} = -\frac{N^2}{(2 \pi)^7 \alpha'^4} \int d^{10}x \sqrt{G}N^\gamma e^{\gamma \phi} \left(R(G) + \beta (\d \phi)^2 - \frac{1}{2(8-p)!N^2} |F_{8-p}|^2 \right).
\end{equation}
This action admits an AdS$_{p+2} \times S^{8-p}$ $(d+1)$-dimensional Euclidean dilaton gravity action given by 
\begin{equation} \label{eq: d+1 dilaton action}
   \mathcal{I}_{d+1} = -\mathcal{N} \int d^{d+1} \sqrt{g}e^{\gamma \phi} \Big( R + \beta (\partial \phi)^2 + C \Big)
\end{equation}
that admits AdS$_{d+1}$ solutions with a $d$-dimensional running scalar
\begin{equation} \label{d1metric}
    ds_{d+1}^2 = \frac{1}{\rho^2} \Big( d\rho^2 + dx dx_D \Big), \quad e^{\phi} = \rho^{2\alpha}
\end{equation}
where \cite{Taylor:2017dly}
\begin{align}\label{symbols}
    &\alpha = -\frac{\gamma}{2(\gamma^2 - \beta)}\\
    &C= \frac{\Big(d(\gamma^2 - \beta) + \gamma^2 \Big) \Big(d(\gamma^2 - \beta) + \beta) \Big)}{(\gamma^2 - \beta)^2}\\
   & \mathcal{N}=\frac{(d_p N)^{(7-p)/(5-p)}g_d^{2(p-3)/(5-p)}R^{(9-p)/(5-p)}}{64 \pi^{(5+p)/2}(2 \pi)^{(p-3)(p-2)/(5-p)} \Gamma\left( \frac{9-p}{2} \right)}.
\end{align}
In the context of M-theory compactifications, the scalar field term plays the role of the warp factor and promoting it to be massive, to account for the lack of observation of this gauge field, is what we refer to as moduli stabilization. 

With a unit AdS radius, the radial coordinate $\rho$ has dimensions of $\text{(length)}^2$ while the dilaton $e^\phi$ has dimensions $\text{(length)}^{2\alpha}$. For a general $\beta$, $d$ and $\gamma$, the equations of motion for the dilation and metric becomes \cite{Kanitscheider:2008kd}
\begin{align}
    &-R_{\mu \nu} + (\gamma^2 - \beta)\d_\mu \phi \d_\nu \phi + \gamma \nabla_\mu \d_\nu \phi - \frac{\gamma^2 + d(\gamma^2 - \beta)}{\gamma^2 - \beta}g_{\mu \nu}=0 \label{eq: eom 1}\\
    & \nabla^2 \phi + \gamma (\d \phi)^2 - \frac{\gamma \left(d(\gamma^2-\beta) + \gamma^2 \right)}{(\gamma^2-\beta)^2}=0 \label{eq: eom 2}\\
    & R+ \beta (\d \phi)^2 + \frac{\left(d(\gamma^2 - \beta) + \gamma^2 \right) \left( d(\gamma^2 - \beta)-\beta \right)}{(\gamma^2 - \beta)^2}\label{eq: eom 3}=0.
\end{align}

The action (\ref{eq: d+1 dilaton action}) can always be written in terms of a reduction of an AdS theory in $(2\sigma + 1)$ dimensions with 
\begin{equation}
2\sigma = (d-2 \alpha \gamma)   
\end{equation}
where $\sigma$ can take non-integer values. 
This reduction is over a $(2\sigma-d)$ torus,
\begin{equation}
    ds_{2\sigma+1}^2 = ds_{d+1}^2+ e^{2\gamma \phi}(dz.dz)_{2\sigma-d}
\end{equation}
where the first term on the RHS is  (\ref{d1metric}) and the second term is the torus metric. For a Dp brane background, the constants are fixed to \cite{Kanitscheider:2008kd}
\begin{align}
    d&=1, \quad 2\alpha \gamma = -1 \\
    \beta &=0, \quad C = d(d+1).
\end{align}
Since we don't have a canonical kinetic term for the scalar field, it can be arbitrarily rescaled, so for $\beta = 0$, we can put $\gamma = 1$.
This gives us NAdS$_2$, from a circle reduction of AdS$_3$ using the Kaluza-Klein ansatz
\begin{equation}
    ds_3^2 = ds_2^2 + e^{2\phi} \Big(dy + A_\mu dx^\mu \Big)^2
\end{equation}
where along with the circle we have included $A_\mu$, an induced Kaluza-Klein gauge field. The scalar field $\phi$ controls the radius of the circle and from here on we will denote the dilaton $e^{\phi}$ as $\Phi$. The pure AdS$_3$ parent action takes the form
\begin{equation}\label{eq: parent action}
    \mathcal{I}_{3}=-\mathcal{N}\int d^3x\sqrt{g}\Phi\left(R+2-\frac{1}{4}\Phi^2 F_{\mu\nu}F^{\mu \nu}\right)  + \mathcal{I}_{\textnormal{ct}}^{(3)}
\end{equation}
where $\mathcal{N}$ is the normalization and where $\mathcal{I}_{\textnormal{ct}}^{(3)}$ are the counterterms including the Gibbson-Hawking-York boundary terms.
$F_{\mu\nu}$ is the Maxwell term from the Kaluza-Klein gauge field.  We will work in the $A_\mu=0$ limit so that the metric does not have any off-diagonal components on the circle. We thus get a simple diagonal uplift$\leftrightarrow$reduction. With the AdS$_3$ metric normalized to have radius one, we get
\begin{equation}
    ds_3^2 = ds_2^2 + \Phi^2 dy^2
\end{equation}
where
\begin{equation}
    ds_2^2=\frac{1}{x^2}\left(-dt^2+dx^2 \right);\quad \Phi=\frac{\phi_r}{x}.
\end{equation}
When reducing AdS$_3$ over a circle, we take the periodicity of the $y$-direction to be $\phi_c$, which is dimensionful and will automatically appear in all thermodynamic quantities. Integrating over the circular direction gives the reduced action of the form
\begin{equation}\label{JTaction}
    \mathcal{I}_2 = -\hat{\mathcal{N}} \int d^2x \sqrt{g} \Phi \left( R+2 \right) +  \Sct
\end{equation}
where $\hat{\mathcal{N}}=\mathcal{N} \phi_c$ \ie $\frac{1}{G_2}=\frac{\phi_c}{G_3}$.
As a boundary condition, we fix the value of the dilaton $\Phi$ at infinity and define the circumference of the circle via the periodicity of $y$. At conformal infinity $\epsilon \to 0$ we thus have
\begin{equation}
    \Phi^2 = \frac{\phi_c^2}{\epsilon^2}.
\end{equation}
All dependence of the periodicity $\phi_c$ will thus be absorbed into an overall prefactor of the action. 

Making the change of coordinates $x^+=t+x$ and $x^-=t-x$, we have 
\begin{equation}
    ds_2^2=\frac{-4dx^+dx^-}{(x^--x^+)^2};\quad \Phi=\frac{2\phi_c}{x^--x^+}
\end{equation}
with a unit curvature length. Now letting $ x^{\pm}=\tanh\frac{\pi y^{\pm}}{\beta}$ we get
\begin{equation}\label{tempmetric}
    ds_2^2 = -\left( \frac{4 \pi^2}{\beta^2} \frac{dy^+ d y^-}{\sinh^2 \left[ \frac{\pi}{\beta}(y^- - y^+)\right]} \right);\quad \Phi=\quad \frac{2\pi\phi_c}{\beta}\frac{1}{\tanh\frac{\pi}{\beta}(y^--y^+)}.
\end{equation}
This is the familiar two-dimensional metric considered in \cite{almheiri2019islands}.  Now, let
\begin{equation}
    y= \sigma + i \tau, \quad \Bar{y} = \sigma - i \tau
\end{equation}
with the Lorenzian time $t=-i \tau$. 
Taken together, we can write the uplifted three-dimensional metric as
\begin{equation} \label{matricsigma}
    ds_3^2 = \left( \frac{2 \pi}{\beta} \right)^2 \left[ \frac{d\sigma^2 + d\tau^2}{\sinh^2\left( \frac{2 \pi \sigma }{\beta}\right)} + \frac{\phi_c^2 dy^2}{\tanh^2\left(\frac{2\pi \sigma}{\beta}\right)} \right], \quad \Phi = -\frac{2 \pi}{\beta} \phi_c \frac{1}{\tanh\left(\frac{2 \pi \sigma}{\beta}\right)}.
\end{equation}
We can also move to the more familiar Schwarzschild coordinates via
\begin{equation}
    \sigma = \left(\frac{\beta}{2\pi} \log \frac{r}{\sqrt{r\left(r+ \frac{4\pi}{\beta}\right)}} \right)
\end{equation}
in which the metric becomes 
\begin{equation} \label{eq: 3d eternal bh}
   ds_3^2 = r \left(r + \frac{4\pi}{\beta} \right)d\tau^2 + \frac{dr^2}{r \left(r+ \frac{4\pi}{\beta}  \right)} + \left( \frac{2\pi \phi_r}{\beta} \right)^2 \left(r \left(r + \frac{4 \pi}{\beta} \right) +1 \right)dy^2.
\end{equation}
This is a metric for an eternal black hole. The horizon is at $r=0$, where the radius of the $y$ circle remains finite. In two dimensions, going between Poincaré to global Schwarzschild coordinates corresponds to going from zero to finite temperature. However, in higher dimensions, this is not the case.

\section{Dimensional analysis}\label{dimanalysis}
We will be studying a cylindrical entangling region with topology $R_t \times R_x \times S^1$ where the circumference of $S^1_y$ is $L_y$.  This cylinder is topologically equivalent to an annulus. 

\noindent
\textbf{Flat Background: }Consider first a CFT$_3$ on the flat background 
\begin{equation}
    ds_3^2 = -dt^2 + dx^2 + \Phi^2 dy^2.
\end{equation}
At zero temperature we have translational invariance in the $x$-direction \ie the entanglement entropy only depends on $\Delta x$ and not on $x_1$ and $x_2$ individually. We also have isometry along the $y$-direction, so the entanglement entropy must be proportional to the circumference $L_y$.  This constrains the zero temperature renormalised entanglement entropy to be of the form 
\begin{equation}\label{dimentropyflat}
\Sren \sim c \frac{L_y}{\Delta x}
\end{equation}
where the dimensionless constant $c$ measures the number of degrees of freedom although it is not exactly a central charge. 

At finite temperature, the renormalised entanglement entropy should take the form
\begin{equation}
\Sren \sim c \frac{L_y}{\Delta x} f\left(\frac{\Delta x}{T} \right)
\end{equation}
where $f\left(\frac{\Delta x}{T} \right)$ is a function of dimensionless ratios which will reduce to (\ref{dimentropyflat}) as $T \to 0$. For large $T$, the expression will asymptote to some power law:
\begin{equation}
\Sren \sim c L_y \frac{T^\alpha}{(\Delta x)^{\alpha +1}}.
\end{equation}

\noindent
\textbf{AdS$_3$ Background: }Now consider the case of a CFT$_3$ on an AdS$_3$ background which in Poincaré coordinates can be written as
\begin{equation}
ds_3^2 = \frac{\ell_3^2}{x^2} \left( -dt^2 + dx^2 + \phi_r^2 dy^2   \right)
\end{equation}
with the conformal factor $\Omega(x)=(\ell_3/x)$. This is a local conformal transformation as it depends on the position $x$ and hence we have now lost the inherent translational invariance so the entanglement entropy generally depends explicitly on $x_1$ and $x_2$\footnote{It might still be the case that there is a translational invariance, but it is not something we can assume from this stage.}. The entangling region geometry has parameters $(L_y, x_1,x_2)$, as well as the implicit parameter $\ell_3$. We still have isometry along the $y$-direction so the entanglement entropy is proportional to $L_y$. Hence, at zero temperature we have 
\begin{equation}\label{dimresult}
\Sren \sim c \frac{L_y}{x_{1,2}} g\left( \frac{\ell_3}{x_2}, \frac{x_1}{x_2} \right)
\end{equation}
where $g\left( \frac{\ell_3}{x_2}, \frac{x_1}{x_2} \right)$ is a function of dimensionless ratios. 

\section{Entanglement entropy on a flat background}\label{sec: flat background}
The renormalised entanglement entropy for a CFT$_3$ living on a flat background was studied in general dimensions in \cite{Taylor:2016aoi} for a disk entangling region. In the case at hand, we will have an annulus entangling region, and our metric is parametrized by $\Phi$ controlling the radius of the circular direction and becomes the dilaton in two dimensions. The metric is given by
\begin{equation}\label{eq: flat 4d metric}
    ds_4^2 = \ell^2 \frac{d\rho^2}{4\rho^2}+\frac{1}{\rho}(-dt^2+dx^2+ \phi_c ^2 dy^2)
\end{equation}
where $\rho$ is an auxiliary bulk direction into which the RT surface propagates. Consider the entangling region of an annulus with width $L$ along the $x$ direction spanning all values of $y$: $0\leq x \leq L$,  $0\leq y \leq 2\pi$, as illustrated in figure {\ref{fig: rt  annulus}}.

\begin{figure}[h]
\centering
\includegraphics[angle=0, width=0.6\linewidth]{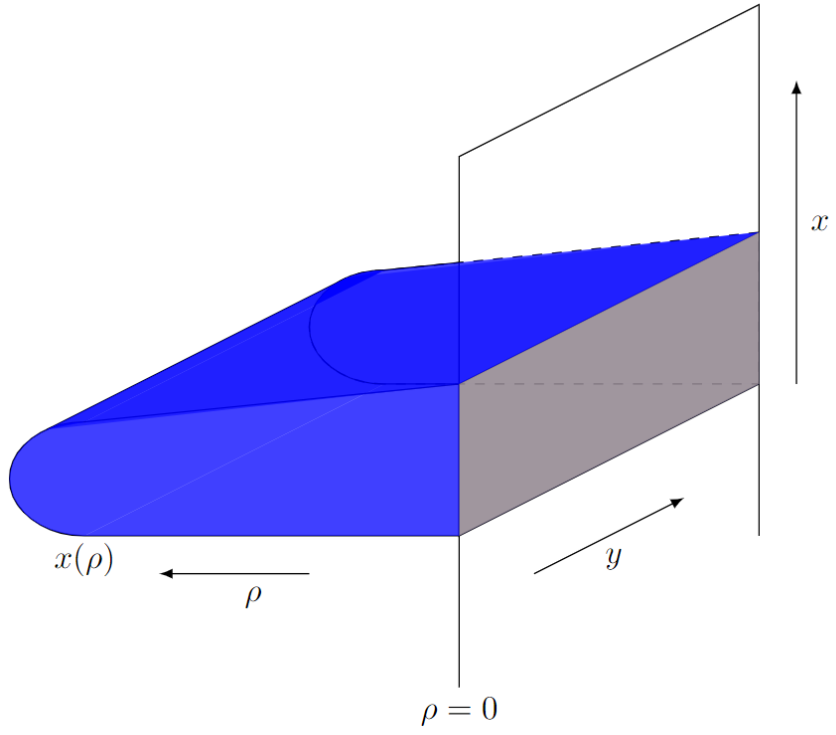}
\caption{A slice of the annulus entangling region (brown), in the circular $y$-direction, at the constant time. The RT surface (blue) propagates into the bulk direction. The equation of motion $x(\rho)$ is the minimal surface minimizing the area functional.}
\label{fig: rt annulus}
\end{figure}

We work with a static gauge in which the time coordinate is constant and the RT surface is spanned by the worldvolume coordinates $ x^\alpha= \{\rho, y\}$, with the AdS$_4$ embedding  $x^m = \{t, \rho, x(\rho), y\}$. The RT surface is symmetric along the $y$ direction due to the isometry along $y$. It is also symmetric about the inflection point $x(\rho_0)=L/2$. When computing the minimal surface, we can thus consider only the area corresponding to $0 \leq x \leq L/2$ and then double it. The induced metric on the RT surface is
\begin{equation} \label{inducedads4}
    h_{\rho \rho} = \frac{\ell^2}{4\rho^2} + \frac{1}{\rho}(\partial_\rho x)^2 \, ,\quad
    h_{yy} = \frac{\phi_c^2}{\rho}.
\end{equation}
The area functional to be minimized becomes  
\begin{equation} \label{areafunctional2}
      \Sreg=\frac{1}{4G_4}\int_0^{2\pi} dy \times 2\int_\epsilon^{\rho_0}d\rho \sqrt{h}= \frac{1}{2G_4}\int_0^{2\pi}dy \int_\epsilon^{\rho_0}d\rho\sqrt{\frac{\phi_c^2 x'(\rho )^2}{\rho ^2}+\frac{\ell^2\phi_c^2}{4 \rho ^3}}
\end{equation}
where $\epsilon$ is a UV-cutoff and the factor of 2 in front of the $\rho$-integral is due to the symmetry around the inflection point. 
Now let us define 
\begin{equation}
    \mathcal{L} = \sqrt{\frac{\phi_c^2 x'(\rho )^2}{\rho ^2}+\frac{\ell^2\phi_c^2}{4 \rho ^3}}.
\end{equation}
Since there is no explicit $x(\rho)$ dependence the equation of motion simply takes the form 
\begin{equation}
\frac{d}{d\rho}\left(\frac{\partial \mathcal{L} }{\partial x'(\rho)}\right) = 0
\end{equation}
which gives us
\begin{equation}
    \frac{\phi_c x'( \rho)}{\rho^2 \sqrt{ \frac{\ell^2}{4\rho^3}+\frac{x'(\rho)^2}{\rho^2} }} = k
\end{equation}
where $k$ is the integration constant.
Solving for $x'(\rho)^2$ gives
\begin{equation} \label{xprime}
 x'(\rho) ^2 = \frac{k^2 \ell^2 \rho}{4 \left( \phi_c^2 - k^2\rho^2  \right)}.
 \end{equation}
 At the turning point, $\rho_0$, we have the boundary condition $x'(\rho_0)\to\infty$ which fixes $k$ as
 \begin{equation} \label{constant}
 k=\frac{\phi_c}{\rho_0}.    
 \end{equation}
Solving further for $x(\rho)$ we get
\begin{equation}
x(\rho)=c_1+\frac{\ell \rho ^{3/2} \, _2F_1\left(\frac{1}{2},\frac{3}{4};\frac{7}{4};\frac{\rho ^2}{\rho _0^2}\right)}{3  \rho_0}
\end{equation}
where $c_1$ is an integration constant. Imposing the boundary condition $x(0)=0$ fixes $c_1=0$. Further, imposing the boundary condition $x(\rho_0)=\frac{L}{2}$ gives
\begin{equation} \label{turningpoint}
    \rho_0=\frac{4 L^2 \Gamma \left(\frac{5}{4}\right)^2}{\ell^2 \pi  \Gamma \left(\frac{3}{4}\right)^2}.
\end{equation} 
Now, substituting (\ref{xprime}) along with (\ref{constant}) into (\ref{areafunctional2}) we get  
\begin{equation}
      S_{\textnormal{reg}} = \frac{ \pi \phi_c \ell}{2G_4} \int_{\epsilon}^{\rho_0} \frac{1}{\sqrt{\rho^3 \left(  1 -\left(\frac{\rho}{\rho_0}\right)^2 \right)}} d\rho.
\end{equation}
By letting $x = \frac{\rho}{\rho_0}$ the above integral can be rewritten as
\begin{equation}
  S_{\textnormal{reg}} = \frac{ \pi \phi_c \ell}{2G_4} \int_{\Tilde{\epsilon}}^{1} \frac{1}{\sqrt{\rho_0}} \frac{x^{-\frac{3}{2}}}{\sqrt{1 - x^2}} dx
\end{equation}
which evaluates to 
\begin{align}
    S_{\rm reg}&=-\frac{\pi  \phi_c \ell\,
   _2F_1\left(-\frac{1}{4},\frac{1}{2};\frac{3}{4};x
   ^2\right)}{ G_4 \sqrt{\rho _0 x}}\Bigg|^1_{\epsilon} \\ 
    &= \frac{\pi ^{3/2}\phi_c \ell \Gamma \left(-\frac{1}{4}\right)
   }{4 G_4 \sqrt{\rho _0} \Gamma
   \left(\frac{1}{4}\right)}+\frac{\pi  \phi_c \ell\,
   _2F_1\left(-\frac{1}{4},\frac{1}{2};\frac{3}{4};\
   \epsilon ^2\right)}{ G_4 \sqrt{\rho _0}
   \sqrt{\epsilon }}.
\end{align}
We could also obtain the upper limit by employing the formula 
\begin{equation}\label{eq: bessel identity}
\int_0^1 dx x^{\mu -1} (1-x^\lambda)^{\nu-1} = \frac{B(\frac{\mu}{\lambda}, \nu)}{\lambda}, \quad B(x,y) = \frac{\Gamma(x) \Gamma(y)}{\Gamma(x+y)}.
\end{equation}
When $\epsilon \to 0$, $S_{\textnormal{reg}}$ has a divergent piece which can be removed by adding the local covariant counterterm
\begin{equation}
    S_{\textnormal{ct}}=-\frac{\pi  \phi_c \ell}{ G_4 \sqrt{\rho _0}
   \sqrt{\epsilon }}.
\end{equation}
We will verify that this indeed is the correct counterterm in section \ref{sec: explicit counterterms}. The renormalised entropy thus  takes the form
\begin{equation}\label{flatS}
      S_{\textnormal{ren}}= - \frac{\pi ^2 \Gamma \left(\frac{3}{4}\right)^2}{G_4 \Gamma \left(\frac{1}{4}\right)^2} \frac{2 \phi_c \ell^2}{L}.
\end{equation}

\section{Entanglement entropy on a curved background}\label{sec: EE curved background}
We will start by considering the AdS$_4$ C-metric, from the Plebanski–Demianski family of type $D$ metrics which is a class of exact solutions to the Einstein field equations with many applications, first studied in \cite{Plebanski:1976gy}. It describes two black holes of opposite charge and equal mass, connected by a non-traversable wormhole. A cosmic string passes through the wormhole which pulls the black holes, causing them to accelerate from each other at a constant rate (see also \cite{Griffiths:2006tk} for a further interpretation of the C-metric). We can write down the C-metric as
\begin{equation} \label{cmetric}
    ds^2= \frac{\ell^2}{(\ell+\hat{r}\rho)^2}\left(-H(\hat{r})dt^2+\frac{d\hat{r}^2}{H(\hat{r})}+\hat{r}^2\left(\frac{d\rho^2}{G(\rho)}+G(\rho) d\tilde{y}^2\right)\right)
\end{equation}
where
\begin{align}
     H(\hat{r})&=\frac{\hat{r}^2}{\ell_3^2}+\kappa-\frac{\mu \ell}{\hat{r}},\\
     G(\rho)&=1-\kappa\rho^2-\mu\rho^3.
\end{align}
$\kappa=\pm 1,0$ gives different background geometries and the AdS$_4$ radius given by 
\begin{equation}\label{eq: bh l formula}
    \ell_4=\left(\frac{1}{\ell^2}+\frac{1}{\ell_3^2}\right)^{-1/2}
\end{equation}
via the Brown–Henneaux holographic formula relating the central charge of the boundary theory with the bulk properties of the black hole \cite{Brown:1986nw}.

We set $\mu=0$ and note that with the change of coordinates $(x,\hat{r})\to(\sigma,r)$ 
\begin{equation}
    \cosh\sigma=\frac{\ell_3}{\ell_4}\frac{\sqrt{1+\frac{\hat{r}^2x^2}{\ell_3^2}}}{\left|1+\frac{\hat{r}x}{l}\right|}, \quad r=\hat{r}\sqrt{\frac{1-\kappa x^2}{1+\frac{\hat{r}^2x^2}{\ell_3^2}}}
\end{equation}
the geometry can be written as
\begin{equation}\label{eq: ads4}
    ds_4^2 = \ell_4^2 d\sigma^2 + \frac{\ell_4^2}{\ell_3^2}\cosh^2\sigma \left(  \frac{dr^2}{\frac{r^2}{\ell_3^2} + \kappa } - \left(\frac{r^2}{\ell_3^2} + \kappa \right)dt^2  + \phi_c^2d\tilde{y}^2\right).
\end{equation}
The conformal boundary is at $\sigma \to \infty$ and the boundary metric is AdS$_3$ expressed in global coordinates \ie
\begin{equation}
    ds_3^2 =    \frac{\ell_4^2}{\ell_3^2} \left( \frac{dr^2}{\frac{r^2}{\ell_3^2} + \kappa } - \left(\frac{r^2}{\ell_3^2} + \kappa \right)dt^2  + \phi_c^2 d\tilde{y}^2 \right).
\end{equation} 
On transforming the conformal AdS$_3$ boundary from global to Poincaré coordinates (see appendix \ref{app: coordinate trans}) we have
\begin{equation}\label{Poincarémet}
    ds_4^2 = d\sigma^2 \ell_4^2 + \ell_4^2 \cosh ^2\sigma  \left( \frac{dx^2-dt^2}{x^2} + \frac{\phi_c^2 dy^2 }{x^2}\right).
\end{equation}
The boundary metric (at $\sigma \to \infty$) is the uplifted AdS$_2$ metric we have been considering:
\begin{equation}\label{eq:Poincaré boundary}
     ds_3^2 = \ell_4^2 \left( \frac{dx^2-dt^2}{x^2} + \frac{\phi_c^2 dy^2 }{x^2} \right).
\end{equation}

Next, we let the RT surface of the annulus be parametrized by the worldvolume coordinates $x^\alpha =\{\sigma, y\}$ and the embedding coordinates are  $x^m=\{t, \sigma, x(\sigma), y\}$.
This gives the area functional for the regulated entropy as 
\begin{equation}\label{Sxsigma}
    S_{\textnormal{reg}}= \frac{1}{4 G_4}\int_0^{2 \pi} dy \left( \int_{\frac{1}{\epsilon}}^{\sigma_0}  d \sigma   \mathcal{L}\left((x_{b}(\sigma), x_{b}'(\sigma), \sigma \right) +   \int_{\sigma_0}^{\frac{1}{\epsilon}} d \sigma    \mathcal{L}\left((x_{a}(\sigma), x_{a}'(\sigma), \sigma \right) \right)
\end{equation}
where 
\begin{equation}
   \mathcal{L}\left((x(\sigma), x'(\sigma), \sigma \right)= \frac{\ell_4^2 \phi_c \cosh \sigma}{x(\sigma )}\sqrt{\frac{\cosh \sigma x'(\sigma
   )^2}{x(\sigma )^2}+1}.
\end{equation}
Here the RT surface does not enjoy a reflection symmetry about the inflection point and we thus get two branches of the solution for the equation of motions: $x_a(\sigma)$, $x_b(\sigma)$.
However, solving the differential equation to obtain the equation of motion will be a daunting task. In appendix \ref{sec: analysis of RT surface} we analyze the RT surface analytically and present an implicit solution. It can be seen that the turning point of the RT surface is close to the conformal boundary. We also perform asymptotic analysis near the conformal boundary and obtain the divergent pieces to the regulated entropy which exactly coincide with the explicit computation of the covariant counterterms in section \ref{sec: explicit counterterms}, as expected. 
To bypass the difficulty of obtaining explicit equations of motions from the RT surface, we will in section \ref{sec: flat limit} present a method to recover the entanglement entropy of a CFT on an AdS$_3$ background from the flat limit.

\subsection{Explicit computation of covariant counterterms}\label{sec: explicit counterterms}
In this section, we explicitly compute the covariant counterterms of the entanglement entropy, using holographic renormalization. Similarly to the disk entangling region considered in \cite{Taylor:2016aoi},  the divergence for an annulus entangling region is manifestly removed by the covariant counterterm 
\begin{equation}
  S_{\textnormal{ct}}= -\frac{1}{4G_4} \int_{\partial A} d^{2}x \sqrt{\Tilde{h}}
\end{equation}
where $\partial A$ is the boundary of the entangling region and $\Tilde{h}$ is the determinant of the induced metric on $\partial A$. We have two disconnected circles at the boundary of the entangling region, one at $x_1$ and the other one at $x_2$.
The embedding coordinates are $x^m = \{t, y,x \}$ and $\partial A$ simply has one coordinate: $x^\alpha = \{ y\}$.
Thus, the induced metric takes the form
\begin{equation}
    \Tilde{h}_{yy} = \left( \partial_y (y) \right)^2 g_{yy} = \ell_4^2 \cosh^2\sigma \frac{\phi_c^2}{x^2}.
\end{equation}
The covariant counterterm thus becomes
\begin{equation} \label{eq: counterterm 3d}
   S_{\textnormal{ct}} = -\frac{1}{4  G_4} \int_{\partial \Sigma}  \ell_4 \cosh\sigma \frac{\phi_c}{x} dy = -\frac{\ell_4 \pi}{4G_4} \frac{\phi_c }{\sqrt{\epsilon}}  \left(\frac{1}{x_2} - \frac{1}{x_1}\right)
\end{equation}
where we in the last line have used that $\lim_{\sigma \to \infty} \cosh\sigma \to \frac{1}{2\sqrt{\epsilon}}$\footnote{it might seem peculiar that the cut-off here is dimensionless but this is because we have used  $\epsilon=u=e^{-2 \sigma}$ so for physical purposes dimensionality for the regulator can easily be reinstated. }. This counterterm exactly agrees with the divergent piece in the asymptotic analysis (\ref{Sdivergent}). The minus sign in the parentheses on the RHS comes because the circle intersecting $x_1$ ($x_2$) is integrated counterclockwise (clockwise). In appendix \ref{sec: analysis of RT surface} we obtain this result from the asymptotic analysis of the area functional (\ref{Sxsigma}).

\subsection{Entanglement entropy from the flat limit}\label{sec: flat limit}
Here we will recover the full explicit solution of the entanglement entropy on a curved background by studying the flat limit. We have from (\ref{Sxsigma}) that the area functional takes the form
\begin{equation}
     \Sreg= \frac{\ell_4^2}{4G_4} (2 \pi \phi_c)\bigg[\int_{\frac{1}{\epsilon}}^{\sigma_0} d \sigma   \mathcal{L}\left((x_{b}(\sigma), x_{b}'(\sigma), \sigma \right) +   \int_{\sigma_0}^{\frac{1}{\epsilon}} d \sigma    \mathcal{L}\left((x_{a}(\sigma), x_{a}'(\sigma), \sigma \right)\bigg] 
\end{equation}
where 
\begin{equation}
\mathcal{L}(x(\sigma),x'(\sigma),\sigma)=\sqrt{\frac{ \cosh ^2\sigma \left(\frac{\cosh^2\sigma x'(\sigma
   )^2}{x(\sigma )^2}+1\right)}{x(\sigma )^2}}.
\end{equation}
By adding the counterterm (\ref{eq: counterterm 3d}) the renormalised entropy becomes
\begin{multline}\label{eq: ren entropy functional}
S_{\textnormal{ren}}=\frac{\ell_4^2}{4G_4} (2\pi \phi_c)\underset{\epsilon \to 0}{\text{lim}}\Bigg[\left(\int_{\frac{1}{\epsilon}}^{\sigma_0} d \sigma   \mathcal{L}|_{x_b(\sigma)} +  \int_{\sigma_0}^{\frac{1}{\epsilon}} d \sigma    \mathcal{L}|_{x_a(\sigma)}\right) -\frac{1}{2\sqrt{\epsilon}}\left(\frac{1}{x_{1}}-\frac{1}{x_{2}}\right)\Bigg].
\end{multline}
From the dimensional analysis, we had that the renormalised entanglement entropy should take the form
\begin{equation}
\Sren \sim c \frac{L_y}{x_1} g\left( \frac{\ell_3}{x_2}, \frac{x_1}{x_2} \right).
\end{equation}
Comparing with (\ref{eq: ren entropy functional}) we make the identifications
\begin{align}
    &c =  \frac{\ell_4^2}{4G_4} \\
    &L_y = 2\pi \phi_c \\ 
    & \frac{1}{x_1} g\left(\frac{\ell_3}{x_{2}}, \frac{x_{1}}{x_{2}} \right)= \underset{\epsilon \to 0}{\text{lim}}\bigg[\Sreg
    -\frac{1}{2\sqrt{\epsilon}}\left(\frac{1}{x_{1}}-\frac{1}{x_{2}}\right)\bigg]\label{g}.
\end{align}
We notice that $g\left(\frac{x_{1}}{x{2}}\right)$ has no explicit $\ell_3$ dependence in Poincaré coordinates, which allows us to recover the renormalised entanglement entropy of a CFT$_3$ on an AdS$_3$ background from the flat limit. 

We can make the above point more explicit by transforming the bulk coordinate in the metric (\ref{Poincarémet})  with $\sigma=\frac{1}{2}\log\rho$,  $\rho \in \left[1, \infty\right)$ which gives
\begin{equation}\label{eq: 4d metric conformally flat}
    ds_4^2=\ell_4^2 \frac{d\rho^2}{4\rho^2} + \ell_4^2 \frac{f(\rho)^2}{x^2} \left(-dt^2 + dx^2 + \phi_c^2 dy^2 \right)
\end{equation}
where
\begin{equation}
f(\rho)^2=\frac{(1+\rho)^2}{4 \rho}.
\end{equation}
The metric (\ref{eq: 4d metric conformally flat}) now looks conformally flat as $\rho \to \infty$. Since the curvature radius $\ell_3$ is only an implicit parameter, the Poincaré coordinates should reduce to flat coordinates in the flat limit $\ell_3 \to \infty$:  $x \to x_f, t\to t_f,y\to y_f$ and $\ell_4 \to \ell$.
Up to the conformal-like factor $\frac{f(\rho)^2}{x^2}$, the metric (\ref{eq: 4d metric conformally flat}) takes the same form as the flat metric (\ref{eq: flat 4d metric}) that was used in calculating the entanglement entropy of a CFT on a flat background. Consequently, the entanglement entropy on a curved background must have the same functional form as the flat entanglement entropy. In other words, the function $g$ in (\ref{dimresult}) must reduce to
\begin{equation}\label{flalim}
\frac{1}{x_{1}}g\left( \frac{x_{1}}{x_{2}} \right) \xrightarrow{\ell_3 \to \infty} \frac{N}{x_{f1}} \left(\frac{ \frac{x_{f1}}{x_{f2}}}{1-\frac{x_{f1}}{x_{f2}}} \right) = \frac{N}{(x_{f2} - x_{f1})}
\end{equation}
where $N$ is a numerical prefactor which we recover from the flat limit:  $N= -2 \pi \frac{ \Gamma \left(\frac{3}{4}\right)^2}{ \Gamma \left(\frac{1}{4}\right)^2}$.
The renormalised entanglement entropy on an AdS$_3$ background then takes the form
\begin{equation}
\Sren^{\text{AdS}_3} = -2 \pi \frac{ \Gamma \left(\frac{3}{4}\right)^2}{ \Gamma \left(\frac{1}{4}\right)^2} c L_y \frac{1}{(x_{2} - x_{1})}.
\end{equation}

\bigskip
\noindent
\textbf{Remarks on conformal-like transformations and the curved boundary of AdS$_4$}
When placing a non-compact space on a conformal boundary one in general has to be careful. The AdS$_4$ C-metric is plagued with conical singularities in the deep exterior owed to the string threading through the wormhole connecting the two black holes. An entangling region can always be sufficiently small so that it is not in casual contact with these conical singularities from the cosmic string. We are only interested in the parametric behavior of the minimal RT surface and only the coordinates of the AdS$_3$ boundary dictates if it is in casual contact with the singularity.  In the context of the island rule when extremizing the entangling region it could in principle be that its minimal RT surface comes in casual contact with the singularities of the bulk manifold. It is not well understood what the implications of this would be and we leave this for future work. In either case, such a discrepancy would be expected to manifest itself in the calculation.

Note that we in general have done an  "illegal" conformal transformation in (\ref{eq: 4d metric conformally flat}) with a long-distance scaling, which is why we call the overall pre-factor to the boundary metric 'conformal-like'.  When going to Poincaré coordinates in (\ref{Poincarémet}) we have done hyperbolic slicing in such a way that only the Poincaré patch of the conformal boundary is considered. All the zero-temperature states live in this patch and since the parent theory is inherently conformal\footnote{not to be confused with the 3$d$ fields that enjoy generalized conformal structure.} the bulk curvature radius would only appear in an overall factor to the entropy; and so, we may use this conformal-like transformation in the case at hand.

\subsection{Area terms}
So far we have studied the contribution from the conformal fields to the entanglement entropy, on a flat and curved background. However, we also have geometrical contributions to the entanglement entropy which we will compute for the respective cases for our annulus entangling region. 
\newline
\noindent
\textbf{Flat case:}
Consider again the flat metric
\begin{equation}
ds_3^2= -dt_f^2+dx_f^2+\phi_c^2dy^2
\end{equation}
where as usual  $0<y\leq 2\pi$. 
If we now consider a surface at point $x$ taking all values of $y$, the induced metric on the boundary of the entangling region, $\d \Sigma$, is $ds_1^2=\phi_c^2 dy^2$. The area of this surface is
\begin{equation}
    \Sarea= \frac{1}{4G_3}\int_{0}^{2\pi}dy \sqrt{\phi_c^2} = \frac{\pi \phi_c}{2 G_3}.
\end{equation}
\newline
\noindent
\textbf{AdS$_3$ case:} With the metric in Poincaré coordinates
\begin{equation}
ds_3^2= \frac{\ell_3^2}{x^2}(-dt^2+dx^2+\phi_c^2dy^2)
\end{equation}
the induced metric on the boundary of the entangling region is $ds_1^2=\left(\frac{\phi_c \ell_3}{x}\right)^2 dy^2$. The area term in this case becomes
\begin{equation} \label{areaterm1}
\Sarea= \frac{1}{4G_3}\int_{0}^{2\pi}dy \sqrt{\frac{\phi_c^2 \ell_3^2}{x^2}} = \frac{\pi \phi_c \ell_3}{2 G_3 x}.
\end{equation}

\section{Generalization to higher dimensions}

\subsection{$d$-dimensional static uncharged black holes in AdS}
We can generalize our analysis to higher dimensions by considering a class of static black hole solutions to Einstein’s equations in $d$ dimensions with a negative cosmological constant and a $(d-2)-$ dimensional horizon topology of positive, zero or negative curvature \cite{Birmingham:1998nr,Mann:1997iz, Banados:1997df}
\begin{align}\label{bhgenmetric}
    ds_d^2 &= -f(r)dt^2 + f^{-1}(r)dr^2 + r^2 h_{ij}(y)dy^i dy^j\\
    f(r) &= k - \frac{\omega_d m}{r^{d-3}}+\frac{r^2}{l^2} \label{f}
\end{align}
with the coordinates labelled as $x^\mu =\{t, r, y^i \}$, $(i=1,...,(d-2))$ and $h_{ij}(y^i)$ being the horizon metric. The horizon is taken to be a compact orientable manifold $M^{d-2}$ with $Vol(M^{d-2})=\int d^{d-2}x\sqrt{h}$, and 
\begin{equation}
    \omega_d = \frac{16 \pi G}{(d-2)Vol(M^{d-2})}.
\end{equation}
$l$ is a parameter with dimensions of length , and $m$ has dimensions of inverse length.
With this form of $f$, one can check that the metric in (\ref{bhgenmetric}) satisfies
\begin{equation}
    R_{\mu\nu}= - \frac{(d-1)}{l^2}g_{\mu\nu}
\end{equation}
with the horizon metric satisfying
\begin{equation}
    R_{ij}(h)= (d-3) k  h_{ij}.
\end{equation}
As pointed out in \cite{Birmingham:1998nr}, the metric (\ref{bhgenmetric}) solves Einstein's field equations for any value of $k$\footnote{as long as the horizon metric is Einstein, it could have positive, negative or zero curvature.} and black hole solutions restricted to horizons with constant curvature are asymptotically locally AdS for all values of $m$.

\subsection{Entangling region in higher dimensions} \label{sec: higher d}
\begin{figure}[h]
\centering
\includegraphics[angle=0, width=0.76\linewidth]{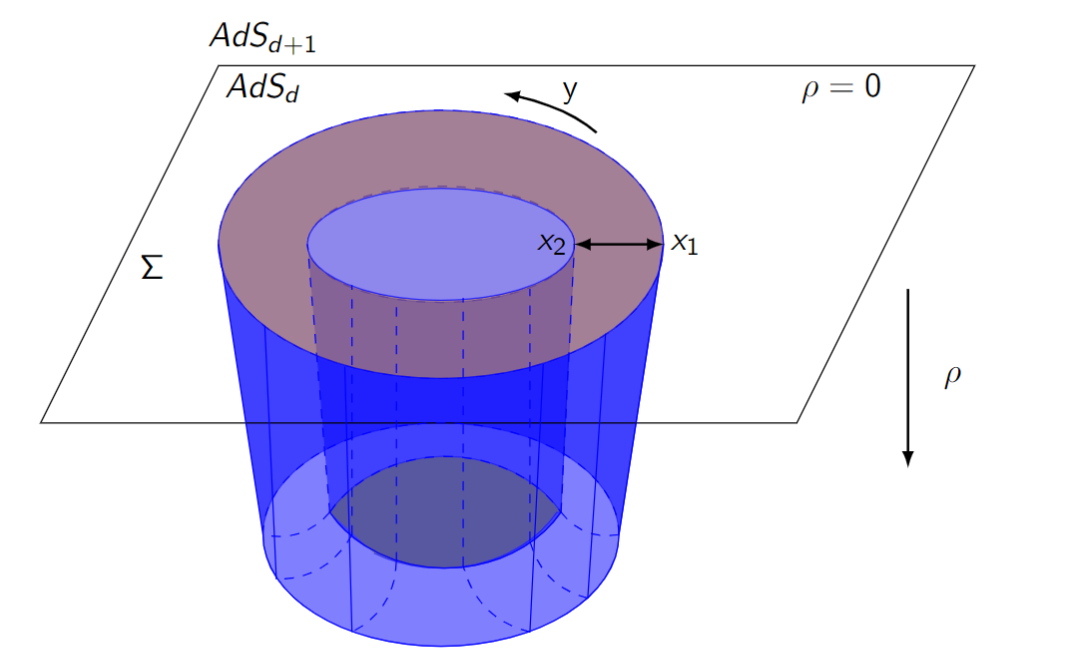}
\caption{Annulus entangling region (brown) at the constant time Cauchy slice $\Sigma$. The RT surface (blue) propagates into the bulk direction. Here $y$ captures the $(d-2)$ isometric direction. }
\label{fig: rt general d}
\end{figure}

We will consider a higher dimensional version of an annulus spanning all values of the $(d-1)$ isometric directions, instead of just one. Thus, the entanglement entropy will still only functionally depend on one coordinate. At a constant time Cauchy slice, our entangling region is a $d$-dimensional manifold, $A=M^{d-2}\times [r_1, r_2]$, as depicted in figure \ref{fig: rt general d}, with all directions in the submanifold $M^{d-2}$ are isometric.
If we did not have complete isometry along the directions of $M^{d-2}$, we would have to consider our entangling region to also depend on the finite length in the corresponding $y^i$ directions. 

\subsection{Structure of the area term in general dimensions}
The geometrical contribution to the generalized entropy is given by the area of the QES \ie the boundary of the entangling region, $\d A$:
\begin{equation}\label{areagen}
    \Sarea = \frac{1}{4 G_{d}} \int_{\d A} d^{d-2} x \sqrt{\tilde{h}}
\end{equation}
where $\tilde{h}_{ij}$ is the induced metric on $\d A$. We consider the black hole solutions in the previous section given by the metric ansatz (\ref{bhgenmetric}) so
\begin{equation}
\tilde{h}_{ij} = r^2 h_{ij} .
\end{equation}
Evaluating the area term (\ref{areagen}) gives
\begin{equation}\label{Sarea}
    \Sarea = \frac{1}{4 G_{d}}\int_{\d A} d^{d-2} x \sqrt{r^{2(d-2)}h} = \frac{\Omega_{d-2}}{4 G_{d}}r^{d-2} ; \quad d>2
\end{equation}
where $ \Omega_{d-2}$ is the volume of the compact orientable manifold capturing the horizon \ie an Einstein space with an arbitrary constant curvature. We note that the area entropy term is monotonic in $r$.

In two dimensions, the boundary of the entangling region is a point and the entangling region is a one-dimensional interval. In three dimensions, the entangling region is circularly uplifted to an annulus, the topology is that of a circle times an interval: $S^1 \times I$, with the QES being the circle. In this case, the extra dimension is an angular coordinate running between $[0,2 \pi)$, so here we would have $\Omega = 2 \pi$.  Since the circle is intrinsically flat we will only get one type of horizon in this case. However, for $d>3$ we could have spherical, toroidal (of genus $\geq$ 1) or hyperbolic horizons.

\subsection{Flat background}
We start with a $(d+1)$-dimensional flat metric
\begin{equation}
    ds^2_{d+1}= \ell_{d+1}^2 \frac{d \rho^2}{4\rho^2} + \frac{1}{\rho}\left( -dt^2 + dr^2   + \phi_c^2 d\Omega_{d-2}^2\right).
\end{equation}
The regulated entanglement entropy becomes
\begin{equation} \label{eq: reg flat d}
    \Sreg = \frac{1}{4G_{d+1}}\Omega_{d-2} \int d\rho \sqrt{\frac{l^2_{d+1}}{4\rho^2} + \frac{r'(\rho)^2}{\rho}}\left( \frac{\phi_c^2}{\rho} \right)^{\frac{d-2}{2}}.
\end{equation}
Next, the equation of motion for the minimal surface is given by
\begin{equation}\label{eq: rprime eom}
    r'(\rho)^2= \frac{\ell_{d+1}}{4 \rho\left(\frac{\rho^{1-d}\phi_c^{2d-4}}{k^2}-1 \right)}
\end{equation}
where $k$ is an integration constant. Using that $r'(\rho) \to \infty$ as $\rho \to \rho_0$ fixes $k$:
\begin{equation}\label{int-d}
    k^2= \phi_c^{2(d-2)} \rho_0^{1-d}.
\end{equation}
The solution for $r(\rho)$ is
\begin{equation}
 r(\rho)=c_1+   \frac{i k \rho^{\frac{d+1}{2}}\phi_c^2 \sqrt{1-k^2 \rho^{d-1}\phi_c^{4-2d}}}{d \sqrt{k^2 \rho^d \phi_c^4 - \rho \phi_c^{2d}}} ._2F_1 \Bigg[ \frac{1}{2}, \frac{d}{2(d-1)}, 1+ \frac{d}{2(d-1)}, k^2 \rho^{d-1} \phi_c^{4-2d} \Bigg]
\end{equation}
where similarly to before, we get that $c_1=0$ from the boundary condition $r(0)=0$. Imposing $r(\rho_0)=\frac{L}{2}$ gives
\begin{equation}\label{eq: turning d}
  \rho_0=\frac{L^2}{\ell_{d+1}^2} \frac{\Gamma\left[\frac{1}{2(d-1)}\right]^2}{\Gamma\left[\frac{d}{2(d-1)}\right]^2 }. 
\end{equation}

Substituting (\ref{eq: rprime eom}) and (\ref{int-d}) back into the integrand of the area functional (\ref{eq: reg flat d}) gives
\begin{equation} \label{eq: reg flat d 2}
    \Sreg= \frac{1}{4G_{d+1}} \Omega_{d-2} \int  \frac{1}{2} \ell_{d+1}\phi_c^{d-2} \frac{1}{\rho^{d/2}} \sqrt{\frac{\rho}{\rho-\rho^d \rho_0^2}}.
\end{equation}
This elliptic integral can be evaluated analytically using the identity
\begin{multline}
    \int \frac{1}{\sqrt{\omega^3(a^2-\omega^2)}} d\omega =2 \frac{2}{a^2 \sqrt{\omega}} \sqrt{a^2-\omega^2}\\
    + \frac{2}{\sqrt{a^3}} \left(F \left(\sin^{-1}\left( \frac{\omega}{a}\right) \big\rvert -1\right) -E \left( \sin^{-1} 
\left(\frac{\omega}{a}\right) \big\rvert  \right)\right)
\end{multline}
where $F(\phi | k^2)$ and $E(\phi|k^2)$ are incomplete elliptic integrals of the first and second kinds, respectively. We get that (\ref{eq: reg flat d 2}) evaluates to 
\begin{equation}\label{eq: s reg flat d}
    \Sreg = \frac{\ell_{d+1}}{3-2d}\sqrt{1-\frac{\rho^{1-d}}{\rho_0^2}} \sqrt{\frac{\rho^{d-1}}{\rho-\rho_0 \rho^d}}\phi_c^d \, _2F_1 \Bigg[\frac{1}{2}, \frac{3-2d}{2-2d}, \frac{5-4d}{2-2d}, \frac{\rho^{1-d}}{\rho_0^2} \Bigg].
\end{equation}
Next, using (\ref{eq: bessel identity}) the renormalised entropy becomes
\begin{equation}\label{eq: Sren d flat}
\begin{align}
    \Sren &= \frac{\ell_{d+1} \phi_c^{d-2}\Omega_{d-2}}{4G_{d+1}} \rho_0^{1-\frac{d}{2}} \frac{\Gamma\left[ \frac{2-d}{d(d-1)}\right] \Gamma \left[ \frac{1}{2}\right]}{\Gamma \left[ \frac{1}{d(d-1)}\right]}\\
    &=\frac{\ell_{d+1}^{d-1} \phi_c^{d-2} \Omega_{d-2}}{4 G_{d+1}}L^{2-d} \frac{2^{d-2}\pi^{\frac{1}{2}(d-1)}\Gamma \left[ \frac{1}{(d-1)} -\frac{2}{d}\right]}{\Gamma \left[ \frac{1}{d(d-1)}\right]} \left(\frac{\Gamma \left[ \frac{1}{2(d-1)}\right]^2}{\Gamma \left[ \frac{d}{2(d-1)}\right]^2}\right)^{1-\frac{d}{2}}.
 \end{align}
\end{equation}
In the last line we substituted the turning point (\ref{eq: turning d}) where the divergence was removed with the $(d+1)-$dimensional counterterm 
\begin{equation}\label{eq: counterterm d+1}
    \Sct = \frac{1}{4G_{d+1}} \int_{\d A} d^{d-1}x \sqrt{h}.
\end{equation}

\subsection{Curved background}
The AdS$_{d+1}$ metric with a boundary covering the Poincaré patch is given by
\begin{equation}
 ds^2=   d \sigma^2 \ell_{d+1}^2 + \ell_{d+1}^2 \cosh^2 \sigma \frac{1}{x^2} (-dt^2 + dx^2 + \phi_c^2 d \Omega_{d-2}).
\end{equation}
Here, the area functional becomes
\begin{equation}
    \Sreg= \frac{\Omega_{d-2}}{4 G_{d+1}} \left( \int_{\frac{1}{\epsilon}}^{\sigma_0}  d \sigma   \mathcal{L}\left((x_{b}(\sigma), x_{b}'(\sigma), \sigma \right) +   \int_{\sigma_0}^{\frac{1}{\epsilon}} d \sigma    \mathcal{L}\left((x_{a}(\sigma), x_{a}'(\sigma), \sigma \right) \right)
\end{equation}
with
\begin{equation}\label{eq: L reg d Poincaré}
   \mathcal{L}= \ell_{d+1}\sqrt{1 + x'(\sigma)^2 \frac{ \cosh^2\sigma}{x(\sigma)^2}}\left( \frac{\ell_{d+1}^2 \cosh^2\sigma \phi_c^2}{x^2} \right)^{\frac{d-2}{2}}.
\end{equation}
By letting $u=e^{-2 \sigma}$ so that the conformal boundary is at $u=0$  we can write the above integrand as
\begin{equation}
\mathcal{L}=\ell_{d+1}^2 \frac{\phi_c}{x(u)} \frac{1}{4u^{3/2}} \sqrt{ (u^2+1)\left( 1+ u(u^2+1) \right)  \frac{x'(u)^2}{x(u)^2}} \left(\ell_{d+1}^2 \frac{(u^2+1)}{4u} \frac{\phi_c^2}{x(u)^2} \right)^{\frac{d-3}{2}}.
\end{equation}
Just as in the three-dimensional example in section \ref{sec: flat limit}, there is no explicit $\ell_d$ dependence, and so, the functional dependence on the non-isometric direction, here the Poincaré coordinate $x$, must be the same as the entropy on the flat background (\ref{eq: Sren d flat}), with the counterterm from (\ref{eq: counterterm d+1}) as
\begin{equation}
    \Sct= -\frac{\Omega_{d-2}}{4 G_{d+1}}\left(\frac{\ell_{d+1}  \phi_c}{ \sqrt{\epsilon}}  \left( \frac{1}{x_2^{d-2}} - \frac{1}{x_1^{d-2}} \right) \right).
\end{equation}

We could also consider the global metric
\begin{equation}
ds^2 = \ell_{d+1}^2 d \sigma^2 + \frac{\ell_{d+1}^2}{\ell_d^2}\cosh^2 \sigma \left( \left(\frac{r^2}{\ell_d^2}+ \kappa \right)dt^2 + \frac{dr^2}{\left(\frac{r^2}{\ell_d^2}+ \kappa \right)} + r^2 \tilde{h}_{ij}\right).
\end{equation}
The regulated entropy now becomes
\begin{equation}\label{eq: s reg d r}
    \Sreg= \frac{\Omega_{d-2}}{4 G_4} \left( \int_{1/\epsilon}^{\sigma_0} \mathcal{L} \big\rvert_{r_1(\sigma)} d\sigma + \int_{\sigma_0}^{1/\epsilon} \mathcal{L} \big\rvert_{r_2(\sigma)} d\sigma \right) 
\end{equation}
where 
\begin{equation}\label{lagrangian}
    \mathcal{L} =\ell_{d+1} \sqrt{ \left(1 + \cosh^2 \sigma \frac{r'(\sigma)^2}{\left(r(\sigma)^2+\kappa \ell_d^2 \right)} \right)} \left(\frac{\ell_{d+1}^2}{\ell_d^2}\cosh^2 \sigma r(\sigma)^2 \right)^{\frac{d-2}{2}}.
\end{equation}
The counterterm here becomes 
\begin{equation}
\Sct=\frac{\Omega_{d-2}}{4G_{d+1}} \left(\frac{\ell_{d+1}}{ \sqrt{\epsilon}\ell_d} \right)^{d-2}  \left( r_2^{d-2} - r_1^{d-2}\right).
\end{equation}

The Poincaré class of solutions corresponds to the $\kappa=0$ class of solutions while letting $y_i \to \ell_d  y_i$. We again find that the explicit $\ell_d$ dependence drops out. 

By again letting $u= e^{-2\sigma}$ as well as $r \to \frac{1}{x}$ in (\ref{eq: s reg d r}) we get
\begin{equation}
    \mathcal{L}= \ell_{d+1}^2\frac{1}{x(u)}\frac{1}{4u^{3/2}} \sqrt{(u^2+1)(1+u(u^2+1))\frac{x'(u)^2}{(x(u)^2 + \kappa \ell_d^2 x(u)^3)}}\left( \ell_{d+1}^2\frac{(u^2+1)}{4u} \frac{1}{x(u)^2}\right)^{\frac{d-3}{2}}.
\end{equation}
It is clear that this reduces to (\ref{eq: L reg d Poincaré}) when $\kappa=0$ up to a factor of $\left( \phi_c\right)^{d-2}$\footnote{ The presence (absence) of $\phi_c$ corresponds to black hole solutions sourced with (without) a running scalar.}.  The entanglement entropy again has the same functional form as the flat result:
\begin{equation}\label{eq: Sren global ads d}
    \Sren = \frac{\ell_{d+1} \Omega_{d-2}}{4G_{d+1}} \ell_d^{d-2}\rho_0^{1-\frac{d}{2}} \frac{\Gamma\left[ \frac{2-d}{d(d-1)}\right] \Gamma \left[ \frac{1}{2}\right]}{\Gamma \left[ \frac{1}{d(d-1)}\right]}.
\end{equation}

\section{Islands in higher dimensions}\label{sec: islands UV}
The island rule (\ref{islands1}) can be written as
\begin{equation}
    S_{\textnormal{EE}} = \textnormal{min}  \Biggr[ \underset{r_2}{\textnormal{ext}}\left(\Sgen=\Scft(r_1,r_2)+\Sarea(r_2)\right) \Biggr]
\end{equation}
where $r_1$ is being fixed and we are extremizing over $r_2$ such that $\SEE$ is minimized.

We showed that the entanglement entropy of conformal fields can be evaluated explicitly in any dimension in the $\kappa=0$ class of solutions. This is the entanglement entropy at zero temperature. However, beyond two dimensions the thermal behavior sits differently in the entanglement entropy \ie one cannot move between zero and finite temperature via a coordinate transformation. In the context of islands, we will thus work with the thermal regulated entropy. 

For an entangling region near the UV boundary, the regulated entanglement entropy is dominated by the divergent piece whose structure is the same in the zero and finite temperature case. Working with the divergent piece could allow us to see if there is a bound on the regulator that would admit islands \ie probing the regime of validity of the UV cutoff that admits islands and beyond which new physics may have to be considered. 

To see whether or not there exists an island with an edge located at a finite point: Let $x_2^*$ be an a local minimum of $\Sgen(x_2, x_1=\text{fixed})$.  A small neighborhood test implies
\begin{align}
& \Scft(x_2) > \Sgen(x_2^*)-\Sarea(x_2)\quad \forall \quad x_2 \neq x_2^* \label{eq: requirement 1}\\
& \Scft'(x_2) \lesseqgtr -\Sarea'(x_2), \quad x_2\lesseqgtr x_2^* \label{eq: requirement 2}
\end{align}
with $\Sarea'(x_2)> 0$ for $x_2\geq 0$ due to the monotonicity of $\Sarea$. If $x_2^*$ instead is a local maximum the inequality between $\Scft'$ and $-\Sarea'$ is flipped.

\bigskip
\noindent
\textbf{Poincaré coordinates:} In this case, the generalized entropy $ \Sgen=-\Sct + \Sarea$ becomes
\begin{equation} \label{eq: sgen poincare}
   \Sgen= \frac{\Omega_{d-2}}{4 G_{d+1}}
\left(\frac{ \ell_{d+1}  \phi_c}{\sqrt{\epsilon}}\right)^{d-2} \left( \frac{1}{x_2^{d-2}} - \frac{1}{x_1^{d-2}} \right) + \frac{\Omega_{d-2}}{4 G_d}  \left(\frac{\phi_c \ell_d}{x_2}\right)^{d-2}.
\end{equation}
For a fixed $x_1$ it is clear that (\ref{eq: sgen poincare}) does not have any extremal points in $x_2$ in any parameter space. Thus, the boundary of the island $x_2^*$ can never be at a finite location. 

\noindent
\textbf{Global coordinates:} In this case, we have
\begin{equation}
    \Sgen = -\frac{\Omega_{d-2} }{4G_{d+1}}  \left(\frac{\ell_{d+1}}{\sqrt{\epsilon} \ell_d}\right)^{d-2}   \left( r_2^{d-2} - r_1^{d-2}\right) + \frac{  \Omega_{d-2} }{4G_d }  r_2^{d-2}
\end{equation}
Similarly, $\Sgen(r_2)$ has no extremal points and is minimized at $r_2=0(\infty)$ for $d=\text{even (odd)}$. Thus, no non-trivial islands can form in this case either.  

We may remain agnostic about the existence of an explicit heatbath beyond the transparent boundary conditions. In two-dimensional gravity, an entangling region cannot be consistently defined a (flat) heatbath is necessary. For $d>2$ the conceptual and technical issue of defining an entangling region in a gauge-invariant way would remain with or without the presence of a heatbath as the factorization of the Hilbert space is a local property and does not dependent on what is going on at the boundary or beyond. A potential inconsistency would be expected to manifest itself in the explicit calculation and making contact with it would be interesting on its own.

\section{Discussion}
In this paper, we study islands by placing fields on an AdS$_{d\geq 3}$ black hole background to model an evaporating black hole. The action of this system is obtained by circularly uplifting NAdS$_2$ gravity. In the island literature, it is often the case that the action of the conformal fields (modeling the Hawking radiation) is supplemented to the gravitational action.  Our construction is more general in the sense that the fields are inherited from the parent theory as supposed to coupled to an already dimensionally reduced gravitational theory. The fields in our construction have a scale (they enjoy generalized conformal structure as opposed to being just conformal) and their symmetries are thus compatible with those of curved backgrounds.
If we reduce our higher dimensional construction to $d=2$ we do not reproduce the familiar JT gravity setting coupled to a CFT$_2$ \cite{almheiri2019islands}; our fields will instead be a CFT$_3$ reduced over a circle. 

The presence of islands in \cite{almheiri2019islands} is owed to the boundary of the spatial entangling region in $2d$ being a zero-dimensional surface, and thus the leading UV divergence coincides with the logarithmic term arising from the Weyl anomaly. 
It is this logarithmic piece that drives the island formation in the sense that it provides the generalized entropy with a local minimum, specifying the finite location of the quantum extremal surface. At the same time, it is also this logarithmic piece that obscures if the entropy is regulated or renormalised. We point out that the non-generic characteristics of the entanglement entropy of a 2$d$ CFT obscures how and and if the JT gravity setting should generalize to higher dimensions.

As a first step in the direction of exemplifying a realization of islands in higher dimensions, we considered the thermal entanglement entropy of a CFT$_{d\geq3}$ on an AdS$_{d\geq3}$ black hole background and take the entangling region to be a (higher-dimensional) annulus. The consequence of working with this entangling region is that when we circularly uplift NAdS$_2$ to higher dimensions, the entanglement entropy will still only depend on one scale: the with of the annulus, as the angular directions are isometric. If we place a part of the entangling region near the UV boundary it will be dominated by the divergent piece given by the covariant counterterms we obtained. Using this as the thermal entropy of the fields, we observe that the generalized entropy will never have any extremal points which means that there cannot exist any non-trivial islands.

It is not necessarily the case that the regulated entropy will expose the same features for islands as a renormalised quantity. The non-generic feature of islands naturally poses the question of what field theories will admit islands. In \cite{landgren1} we specify the space of CFTs that will admit islands. The relevant piece to consider that captures the details of the island is the variation of the entanglement entropy under an inhomogeneous transformation of the entangling region \ie a generalization of (\ref{liu}).

\section*{Acknowledgements}
It is a pleasure to thank Marika Taylor for guidance throughout the project and Felix Haehl for feedback on the manuscript. The work of FL was supported by STFC (ST/W507799/1).

\appendix

\section{Analysis of the RT surface}\label{sec: analysis of RT surface}

The area functional (\ref{Sxsigma}) is minimized by solving the differential equation
\begin{multline}
   \cosh (\sigma ) x(\sigma )^2
   \left(\cosh (\sigma ) x''(\sigma )+3 \sinh (\sigma ) x'(\sigma )\right) \\+ 2 \sinh (\sigma ) \cosh ^3(\sigma ) x'(\sigma )^3 +x(\sigma)^3=0.
\end{multline}
The RT surface is the solution $x(\sigma)$ that has a turning point at $(x_0,\sigma_0)$ in the bulk and intersects the boundary at $(\sigma \to \infty,x_1)$ and $(\sigma \to \infty,x_2=x_1+L)$. We expect two branches of solution corresponding to whether the solution intersects the boundary at $x_1$ or $x_2$: $x_a(\sigma)$ and $x_b(\sigma)$. Hence, we have the boundary conditions
\begin{align}
    x_a(\infty)&=x_1, \quad x_b(\infty)=x_2\\
    x_a(\sigma_0)&=x_b(\sigma_0)=x_0\\
    x_a'(\sigma_0)&=x_b'(\sigma_0)=\infty  .
\end{align}
Doing a change of coordinates $u=e^{-2\sigma}$, the differential equation becomes
\begin{multline}\label{xdeqn}
    (u-1) u (u+1)^3 x'(u)^3+\frac{1}{2} (u+1) x(u)^2 \big(2 u (u+1) x''(u)\\+(5 u-1)
   x'(u)\big)+x(u)^3=0
\end{multline}
where the conformal boundary now is at $u=0$. Further changing coordinates to $x(u)=e^{f(u)}$ we get,
\begin{multline}
    u (u+1)^2 f''(u)+\frac{1}{2} \left(5 u^2+4 u-1\right) f'(u)+(u-1) u (u+1)^3
   f'(u)^3 \\+u (u+1)^2 f'(u)^2+1=0.
\end{multline}
We can immediately notice that the resulting differential equation depends only on $f''(u)$ and $f'(u)$. Hence we can now split the second-order ODE into two first-order ODEs:
\begin{align}
    &f'(u)=g(u)\label{fdeqn}\\
   \begin{split} &u (u+1)^2 g'(u)+\frac{1}{2} \left(5 u^2+4 u-1\right) g(u)+(u-1) u (u+1)^3 g(u)^3\\
   &+u(u+1)^2 g(u)^2+1=0. \label{gdeqn}
   \end{split}
\end{align}
Solving (\ref{gdeqn}) we get the implicit relation for $g(u)$
\begin{equation}\label{gimplicit}
    \frac{\sqrt{\frac{u-1}{u}} \left(\frac{(2 (u+1) u g(u)-u+1) \, _2F_1\left(\frac{1}{4},1;\frac{3}{2};-\frac{(-2 (u+1) g(u) u+u-1)^2}{u \left(\left(u^2-1\right)
   g(u)+2\right)^2}\right)}{\left(u^2-1\right) g(u)+2}+u-1\right)}{2 \sqrt{1-u} \sqrt[4]{-\frac{(-2 (u+1) u g(u)+u-1)^2}{u \left(\left(u^2-1\right) g(u)+2\right)^2}-1}}=C_1
\end{equation}
where $C_1$ is the integration constant. Reinstating the coordinates $x(u)$ we have 
\begin{equation}\label{gxrln}
    g(u)=f'(u)=\frac{\partial(\log[x(u)])}{\partial u}=\frac{x'(u)}{x(u)}
\end{equation}
Substituting this back in (\ref{gimplicit}) and imposing the boundary condition at the turning point $x'(u_0=e^{-2\sigma_0})=\infty$ we fix $C_1$ in terms of $u_0=e^{-2\sigma_0}$    :  
\begin{equation}\label{C1}
    C_1(u_0)=-\frac{\sqrt{\frac{u_0-1}{u_0}} \left(2 u_0 \, _2F_1\left(\frac{1}{4},1;\frac{3}{2};-\frac{4 u_0}{(u_0-1)^2}\right)+(u_0-1)^2\right)}{2
   (1-u_0)^{3/2} \sqrt[4]{-\frac{(u_0+1)^2}{(u_0-1)^2}}}
\end{equation}
So $C_1$ encodes the information about only the turning point.
Now considering (\ref{gimplicit}) and (\ref{gxrln}), we have the general relation
\begin{equation}
    g(u)=\frac{x'(u)}{x(u)}= P(u,C_1(u_0)) 
\end{equation}
for a general function $ P(u,C_1(u_0))$. Solving for $x(u)$ gives us
\begin{equation}\label{xgrln}
    x(u)=C_2e^{\int du P(u,C_1)}
\end{equation}
where $C_2$ is the second integration constant that acts as an overall scaling. This can also be observed from the differential equation for $x(u)$ (\ref{xdeqn}) where we see that $C_2 x(u)$ is a solution if $x(u)$ is a solution. Below we will see that the asymptotic analysis of $u\to0$  shows that $e^{\int du P(u,C_1)} \to 1$ as $u\to 0$.

Now, imposing the boundary condition $x(0)=x_1, x_2$ along with $x'(u_0)=\infty$, we get two branches of solutions, one with $C_2=x_1$ and the other with $C_2=x_2$.  $C_2$ is independent of the choice of $C_1$. In other words, $C_2$ only captures where the curve intersects the boundary and is independent of $C_1$ which only captures information about the turning point $u_0$. 

Close to the boundary, we can write down the following ansatz for a particular $g(u)$:
   \begin{equation}
    g(u) = \sum_{n=0}^{\infty}a_n u^n.
\end{equation}
Plugging this ansatz in ($\ref{gdeqn}$) and solving perturbatively order by order for $a_n$ we get
\begin{equation}\label{gpart}
    g(u) = \sum_{n=0}^{\infty}2 u^{2n}=\frac{2}{1-u^2}.
\end{equation}
This is a particular solution for $g(u)$. Reinstating the coordinates $x(u)=e^{\int du g(u)}$ we get a one-parameter family of solutions for $x(u)$
\begin{equation}\label{opx}
    x(u) = C_3 \left(\frac{1+u}{1-u}\right).
\end{equation}
From our previous analysis of the full solution for $x(u)$ we see that this particular solution corresponds to a choice of the integration constant $C_1(u_0)$. $C_3$ in this particular solution is the scaling constant.  Since $C_3$ is independent of $C_1$, we could plug in the derivative of the particular solution for $x(u)$ (\ref{opx}) into (\ref{gxrln}) and (\ref{gimplicit}), to get an implicit full solution for $x(u)$. 
Combining this with the results we got for $C_1(u_0)$ we get,
\begin{multline}
   \frac{\sqrt{\frac{u-1}{u}} \left(\frac{\left(\frac{4 C_3 (u+1) u}{(u-1)^2
   x(u)}-u+1\right) \, _2F_1\left(\frac{1}{4},1;\frac{3}{2};-\frac{\left(-\frac{4 (u+1) C_3 u}{(u-1)^2 x(u)}+u-1\right){}^2}{u \left(\frac{2 \left(u^2-1\right) C_3}{(u-1)^2
   x(u)}+2\right){}^2}\right)}{\frac{2 C_3 \left(u^2-1\right)}{(u-1)^2 x(u)}+2}+u-1\right)}{2 \sqrt{1-u} \sqrt[4]{-\frac{\left(-\frac{4 C_3 (u+1) u}{(u-1)^2 x(u)}+u-1\right){}^2}{u
   \left(\frac{2 C_3 \left(u^2-1\right)}{(u-1)^2 x(u)}+2\right){}^2}-1}}\\=C_1(u_0)= -\frac{\sqrt{\frac{u_0-1}{u_0}} \left(2 u_0 \, _2F_1\left(\frac{1}{4},1;\frac{3}{2};-\frac{4 u_0}{\left(u_0-1\right){}^2}\right)+\left(u_0-1\right){}^2\right)}{2
   \left(1-u_0\right){}^{3/2} \sqrt[4]{-\frac{\left(u_0+1\right){}^2}{\left(u_0-1\right){}^2}}}.
\end{multline}
This implicit solution for $x(u)$ is still difficult to unpack and in the next subsection, we will see how the equation of motions behaves near the boundary.

\subsection{Asymptotic analysis at the boundary}
Consider expanding the particular solution (\ref{gpart}) near the boundary. 
\newline Since $0<u\leq u_0 <1$ a natural expansion parameter for a perturbative series is any function $f(u)$ such that $0<f(u)<1$. We choose the expansion parameter $f(u)=q=\sqrt{u}$ and consider an ansatz for $g(u)$ of the form
\begin{equation}
    g(u)= \frac{2}{1-u^2}+q \sum_{n=0}^{\text{order}}h_n q^n.
\end{equation}
We can plug this ansatz into the differential equation for $g(u)$ (\ref{gdeqn}) and solve for the $h_n$'s order by order perturbatively. We have listed the $h_n$'s up to $h_6$ below: 

\begin{align}
    h_0 &= k,\\
    h_1 &= 0,\\
    h_2 &= 5k,\\
    h_3 &= (10 k^2)/3,\\
    h_4 &= k (28 + k^2)/2,\\
    h_5 &= (80 k^2)/3,\\
    h_6 &= 30 k + (305 k^3)/18
\end{align}
where $k$ is the integration constant. Reinstating the coordinates $x(u)=e^{\int du g(u)}$ we get,
\begin{equation}\label{asymx}
    x(u; k, C_2) = C_2 \frac{1+u}{1-u} e^{\frac{2}{3} k u^{3/2}\left(1+ 3 \sum \limits_{n=2}^{\text{order}} \frac{h_n}{k} u^{n/2} \right)}
\end{equation}
Since $C_2$ is just the scaling constant and $x(u) \to C_2$ as $u\to0$, therefore $C_2 = x_1, x_2$. These two choices along with corresponding choices for the constant $k=k_1,k_2$ gives two branches of solutions, $x_a(u;x_1, k_1)$ and $x_b(u;x_2, k_2)$, on which the matching boundary conditions at the turning point have to be imposed to fix $ k_1(x_1,x_2)$ and $k_2(x_1,x_2)$.

\subsection{Divergent terms in the regulated entropy}

Since the divergent contributions to the entropy functional come from close to the boundary, we can extract this divergence by simply considering the asymptotic series solution of $x(u)$ around the boundary, up to suitable orders that capture the full divergent behavior.

In $x(u)$ coordinates the entropy functional (\ref{Sxsigma}) takes the form
\begin{align}\label{sxu}
    S_{\textnormal{reg}} &= \frac{1}{4G_4}\int _0^{2\pi} dy \left(\int_{\epsilon}^{u_0} d u \mathcal{L}(u, x_a(u;k_1,x_1)) + \int_{u_0}^{\epsilon} du \mathcal{L}(u; x_b(u;k_2,x_2)) \right), \\
     \mathcal{L}(u) &= \frac{-1}{4 u} \sqrt{\frac{\ell_4^4 (u+1)^2 \phi ^2 \left(u (u+1)^2 x'(u)^2+x(u)^2\right)}{u x(u)^4}}
\end{align}
where $x_a(u;k_1,x_1)$ and $x_b(u;k_2,x_2')$ are the two branches intersecting the boundary at $x_1, x_2$ respectively.

Substituting the asymptotic series solution of $x(u; k,C_2)$ around the boundary (\ref{asymx}) into $\mathcal{L}(u)$, and expanding around $u=0$ gives
\begin{equation}
    \mathcal{L}(u;k, C_2) = \frac{\phi \ell_4^2}{C_2} \left( \frac{-1}{4 u^{3/2}}-\frac{1}{4 u^{1/2}}-\frac{k}{3}-\frac{k^2}{8} u^{1/2}-\frac{4y}{3} u-\frac{125 k^2}{72} u^{3/2} \right) + \mathcal{O} (u^2). 
\end{equation}
Only the first term $\frac{\phi \ell_4^2}{C_2} \left( \frac{-1}{4 u^{3/2}}\right)$ in $\mathcal{L}(u;k, C_2)$ contributes to the divergence in the entanglement entropy. Since we are considering the series solution of $x(u)$ around the boundary from where the divergent contributions reside, more terms in the asymptotic series for $x(u)$ will not give additional contributions to the divergence. 

Plugging $\mathcal{L}(u)$ back into the entropy functional (\ref{sxu}), we get the divergent contribution to the entanglement entropy in full generality given by
\begin{equation}\label{Sdivergent}
    S_{\text{div}} =  \frac{\pi\phi \ell_4^2}{4G_4 \sqrt{\epsilon}}\left(\frac{1}{x_2} - \frac{1}{x_1}\right)
\end{equation}

\section{Coordinate transformations }\label{app: coordinate trans}
\textbf{Global coordinates:}
Consider the global metric
\begin{equation}
    ds^2=-\left(1+\frac{r^2}{\ell_3^2}\right) d\tau^2 +\left(1+\frac{r^2}{\ell_3^2}\right)^{-1}dr^2+r^2d\theta^2
\end{equation}
where $-\pi\leq\theta<\pi$.
By putting $r=\ell_3 \sinh(\rho)$ we obtain the metric
\begin{equation}
    ds^2=-\cosh^2(\rho)d\tau^2 + \ell_3^2d\rho^2+\ell_3^2\sinh^2(\rho)d\theta^2
\end{equation}
Now, letting $\rho=\sinh^{-1}(\tan(\Tilde{\rho}/\ell_3))$ and $\theta=\Tilde{\theta}/\ell_3$ we get
\begin{equation}
    ds^2 = \frac{1}{\cos^2\left(\frac{\Tilde{\rho}}{\ell_3}\right)}\left(-d\tau^2+d\Tilde{\rho}^2+\sin^2\left(\frac{\Tilde{\rho}}{\ell_3}\right)d\Tilde{\theta}^2\right).
\end{equation}
\newline
\textbf{Poincaré coordinates:}
The metric in Poincaré coordinates reads
\begin{equation}
 ds^2=\frac{\ell_3^2}{x^2}(-dt^2+dx^2+dy^2)  
\end{equation} 
and can be achieved via the transformations
\begin{align}
\sqrt{\ell_3^2+r^2} \cos{(\ell_3 \tau)} &= \frac{\ell_3 \alpha^2 + \ell_3(-t^2+x^2+y^2)}{2\alpha x},\\
\sqrt{\ell_3^2+r^2}\sin{(\ell_3 \tau)}&=\frac{\ell_3 t}{x_p},\\
r\sin{\theta}&=\frac{\ell_3 y}{x},\\
-r \cos{\theta}&=\frac{-\ell_3 \alpha^2 + \ell_3(-t^2+x^2+y^2)}{2\alpha x}
\end{align} 
where $\alpha$ is an arbitrary real number corresponding to a particular isometry of AdS.

\newpage
\bibliographystyle{JHEP}
\bibliography{bib}

\providecommand{\href}[2]{#2}\begingroup\raggedright\begin{thebibliography}{10}

\bibitem{Casini:2022rlv}
H.~Casini and M.~Huerta, \emph{{Lectures on entanglement in quantum field theory}}, \href{https://doi.org/10.22323/1.403.0002}{\emph{PoS} {\bfseries TASI2021} (2023) 002} [\href{https://arxiv.org/abs/2201.13310}{{\ttfamily 2201.13310}}].

\bibitem{Calabrese:2004eu}
P.~Calabrese and J.~L. Cardy, \emph{{Entanglement entropy and quantum field theory}}, \href{https://doi.org/10.1088/1742-5468/2004/06/P06002}{\emph{J. Stat. Mech.} {\bfseries 0406} (2004) P06002} [\href{https://arxiv.org/abs/hep-th/0405152}{{\ttfamily hep-th/0405152}}].

\bibitem{Cardy:2014jwa}
J.~Cardy and C.~P. Herzog, \emph{{Universal Thermal Corrections to Single Interval Entanglement Entropy for Two Dimensional Conformal Field Theories}}, \href{https://doi.org/10.1103/PhysRevLett.112.171603}{\emph{Phys. Rev. Lett.} {\bfseries 112} (2014) 171603} [\href{https://arxiv.org/abs/1403.0578}{{\ttfamily 1403.0578}}].

\bibitem{Herzog:2014fra}
C.~P. Herzog, \emph{{Universal Thermal Corrections to Entanglement Entropy for Conformal Field Theories on Spheres}}, \href{https://doi.org/10.1007/JHEP10(2014)028}{\emph{JHEP} {\bfseries 10} (2014) 028} [\href{https://arxiv.org/abs/1407.1358}{{\ttfamily 1407.1358}}].

\bibitem{Herzog:2015cxa}
C.~P. Herzog and M.~Spillane, \emph{{Thermal corrections to R\'enyi entropies for free fermions}}, \href{https://doi.org/10.1007/JHEP04(2016)124}{\emph{JHEP} {\bfseries 04} (2016) 124} [\href{https://arxiv.org/abs/1506.06757}{{\ttfamily 1506.06757}}].

\bibitem{Vassilevich:2003xt}
D.~V. Vassilevich, \emph{{Heat kernel expansion: User's manual}}, \href{https://doi.org/10.1016/j.physrep.2003.09.002}{\emph{Phys. Rept.} {\bfseries 388} (2003) 279} [\href{https://arxiv.org/abs/hep-th/0306138}{{\ttfamily hep-th/0306138}}].

\bibitem{penington2020entanglement}
G.~Penington, \emph{{Entanglement Wedge Reconstruction and the Information Paradox}}, \href{https://doi.org/10.1007/JHEP09(2020)002}{\emph{JHEP} {\bfseries 09} (2020) 002} [\href{https://arxiv.org/abs/1905.08255}{{\ttfamily 1905.08255}}].

\bibitem{almheiri2019islands}
A.~Almheiri, R.~Mahajan and J.~Maldacena, \emph{{Islands outside the horizon}},  \href{https://arxiv.org/abs/1910.11077}{{\ttfamily 1910.11077}}.

\bibitem{Casini:2004bw}
H.~Casini and M.~Huerta, \emph{{A Finite entanglement entropy and the c-theorem}}, \href{https://doi.org/10.1016/j.physletb.2004.08.072}{\emph{Phys. Lett. B} {\bfseries 600} (2004) 142} [\href{https://arxiv.org/abs/hep-th/0405111}{{\ttfamily hep-th/0405111}}].

\bibitem{ryu2006aspects}
S.~Ryu and T.~Takayanagi, \emph{{Aspects of Holographic Entanglement Entropy}}, \href{https://doi.org/10.1088/1126-6708/2006/08/045}{\emph{JHEP} {\bfseries 08} (2006) 045} [\href{https://arxiv.org/abs/hep-th/0605073}{{\ttfamily hep-th/0605073}}].

\bibitem{Taylor:2016aoi}
M.~Taylor and W.~Woodhead, \emph{{Renormalized entanglement entropy}}, \href{https://doi.org/10.1007/JHEP08(2016)165}{\emph{JHEP} {\bfseries 08} (2016) 165} [\href{https://arxiv.org/abs/1604.06808}{{\ttfamily 1604.06808}}].

\bibitem{Almheiri:2019psy}
A.~Almheiri, R.~Mahajan and J.~E. Santos, \emph{{Entanglement islands in higher dimensions}}, \href{https://doi.org/10.21468/SciPostPhys.9.1.001}{\emph{SciPost Phys.} {\bfseries 9} (2020) 001} [\href{https://arxiv.org/abs/1911.09666}{{\ttfamily 1911.09666}}].

\bibitem{Chen:2020uac}
H.~Z. Chen, R.~C. Myers, D.~Neuenfeld, I.~A. Reyes and J.~Sandor, \emph{{Quantum Extremal Islands Made Easy, Part I: Entanglement on the Brane}}, \href{https://doi.org/10.1007/JHEP10(2020)166}{\emph{JHEP} {\bfseries 10} (2020) 166} [\href{https://arxiv.org/abs/2006.04851}{{\ttfamily 2006.04851}}].

\bibitem{Chen:2020hmv}
H.~Z. Chen, R.~C. Myers, D.~Neuenfeld, I.~A. Reyes and J.~Sandor, \emph{{Quantum Extremal Islands Made Easy, Part II: Black Holes on the Brane}}, \href{https://doi.org/10.1007/JHEP12(2020)025}{\emph{JHEP} {\bfseries 12} (2020) 025} [\href{https://arxiv.org/abs/2010.00018}{{\ttfamily 2010.00018}}].

\bibitem{Geng:2020qvw}
H.~Geng and A.~Karch, \emph{{Massive islands}}, \href{https://doi.org/10.1007/JHEP09(2020)121}{\emph{JHEP} {\bfseries 09} (2020) 121} [\href{https://arxiv.org/abs/2006.02438}{{\ttfamily 2006.02438}}].

\bibitem{Neuenfeld:2021wbl}
D.~Neuenfeld, \emph{{The Dictionary for Double Holography and Graviton Masses in d Dimensions}},  \href{https://arxiv.org/abs/2104.02801}{{\ttfamily 2104.02801}}.

\bibitem{Randall:1999vf}
L.~Randall and R.~Sundrum, \emph{{An Alternative to compactification}}, \href{https://doi.org/10.1103/PhysRevLett.83.4690}{\emph{Phys. Rev. Lett.} {\bfseries 83} (1999) 4690} [\href{https://arxiv.org/abs/hep-th/9906064}{{\ttfamily hep-th/9906064}}].

\bibitem{Randall:1999ee}
L.~Randall and R.~Sundrum, \emph{{A Large mass hierarchy from a small extra dimension}}, \href{https://doi.org/10.1103/PhysRevLett.83.3370}{\emph{Phys. Rev. Lett.} {\bfseries 83} (1999) 3370} [\href{https://arxiv.org/abs/hep-ph/9905221}{{\ttfamily hep-ph/9905221}}].

\bibitem{Karch:2000ct}
A.~Karch and L.~Randall, \emph{{Locally localized gravity}}, \href{https://doi.org/10.1088/1126-6708/2001/05/008}{\emph{JHEP} {\bfseries 05} (2001) 008} [\href{https://arxiv.org/abs/hep-th/0011156}{{\ttfamily hep-th/0011156}}].

\bibitem{Giddings:2000mu}
S.~B. Giddings, E.~Katz and L.~Randall, \emph{{Linearized gravity in brane backgrounds}}, \href{https://doi.org/10.1088/1126-6708/2000/03/023}{\emph{JHEP} {\bfseries 03} (2000) 023} [\href{https://arxiv.org/abs/hep-th/0002091}{{\ttfamily hep-th/0002091}}].

\bibitem{Karch:2000gx}
A.~Karch and L.~Randall, \emph{{Open and closed string interpretation of SUSY CFT's on branes with boundaries}}, \href{https://doi.org/10.1088/1126-6708/2001/06/063}{\emph{JHEP} {\bfseries 06} (2001) 063} [\href{https://arxiv.org/abs/hep-th/0105132}{{\ttfamily hep-th/0105132}}].

\bibitem{Ryu:2006bv}
S.~Ryu and T.~Takayanagi, \emph{{Holographic derivation of entanglement entropy from AdS/CFT}}, \href{https://doi.org/10.1103/PhysRevLett.96.181602}{\emph{Phys. Rev. Lett.} {\bfseries 96} (2006) 181602} [\href{https://arxiv.org/abs/hep-th/0603001}{{\ttfamily hep-th/0603001}}].

\bibitem{Laflorencie:2015eck}
N.~Laflorencie, \emph{{Quantum entanglement in condensed matter systems}}, \href{https://doi.org/10.1016/j.physrep.2016.06.008}{\emph{Phys. Rept.} {\bfseries 646} (2016) 1} [\href{https://arxiv.org/abs/1512.03388}{{\ttfamily 1512.03388}}].

\bibitem{Hawking_2005}
S.~W. Hawking, \emph{Information loss in black holes}, \href{https://doi.org/10.1103/physrevd.72.084013}{\emph{Physical Review D} {\bfseries 72} (2005) }.

\bibitem{landgren1}
F.~Landgren, A.~Shekar and M.~Taylor{\emph{\quad to appear} (2023) }.

\bibitem{Compere:2013bya}
G.~Comp\`ere, W.~Song and A.~Strominger, \emph{{New Boundary Conditions for AdS3}}, \href{https://doi.org/10.1007/JHEP05(2013)152}{\emph{JHEP} {\bfseries 05} (2013) 152} [\href{https://arxiv.org/abs/1303.2662}{{\ttfamily 1303.2662}}].

\bibitem{Castro:2014ima}
A.~Castro and W.~Song, \emph{{Comments on $\mathrm{AdS}_2$ Gravity}},  \href{https://arxiv.org/abs/1411.1948}{{\ttfamily 1411.1948}}.

\bibitem{Hartman:2008dq}
T.~Hartman and A.~Strominger, \emph{{Central Charge for AdS(2) Quantum Gravity}}, \href{https://doi.org/10.1088/1126-6708/2009/04/026}{\emph{JHEP} {\bfseries 04} (2009) 026} [\href{https://arxiv.org/abs/0803.3621}{{\ttfamily 0803.3621}}].

\bibitem{Strominger:1998yg}
A.~Strominger, \emph{{AdS(2) quantum gravity and string theory}}, \href{https://doi.org/10.1088/1126-6708/1999/01/007}{\emph{JHEP} {\bfseries 01} (1999) 007} [\href{https://arxiv.org/abs/hep-th/9809027}{{\ttfamily hep-th/9809027}}].

\bibitem{Almheiri:2014cka}
A.~Almheiri and J.~Polchinski, \emph{{Models of AdS$_{2}$ backreaction and holography}}, \href{https://doi.org/10.1007/JHEP11(2015)014}{\emph{JHEP} {\bfseries 11} (2015) 014} [\href{https://arxiv.org/abs/1402.6334}{{\ttfamily 1402.6334}}].

\bibitem{Sachdev_1993}
S.~Sachdev and J.~Ye, \emph{Gapless spin-fluid ground state in a random quantum heisenberg magnet}, \href{https://doi.org/10.1103/physrevlett.70.3339}{\emph{Physical Review Letters} {\bfseries 70} (1993) 3339}.

\bibitem{kitaev2015simple}
A.~Kitaev, \emph{A simple model of quantum holography,(2015) http://online. kitp. ucsb. edu/online/entangled15/kitaev/, http://online. kitp. ucsb. edu/online/entangled15/kitaev2/, talks at kitp},  2015.

\bibitem{Jackiw:1982hg}
R.~Jackiw, \emph{{Liouville field theory: a two-dimensional model for gravity?, 1982}}, .

\bibitem{Teitelboim:1983fg}
C.~Teitelboim, \emph{{The Hamiltonian structure of two-dimensional space-time and its relation with the conformal anomaly, 1983}}, .

\bibitem{Mertens:2022irh}
T.~G. Mertens and G.~J. Turiaci, \emph{{Solvable Models of Quantum Black Holes: A Review on Jackiw-Teitelboim Gravity}},  \href{https://arxiv.org/abs/2210.10846}{{\ttfamily 2210.10846}}.

\bibitem{Kanitscheider:2008kd}
I.~Kanitscheider, K.~Skenderis and M.~Taylor, \emph{{Precision holography for non-conformal branes}}, \href{https://doi.org/10.1088/1126-6708/2008/09/094}{\emph{JHEP} {\bfseries 09} (2008) 094} [\href{https://arxiv.org/abs/0807.3324}{{\ttfamily 0807.3324}}].

\bibitem{Boonstra:1998mp}
H.~J. Boonstra, K.~Skenderis and P.~K. Townsend, \emph{{The domain wall / QFT correspondence}}, \href{https://doi.org/10.1088/1126-6708/1999/01/003}{\emph{JHEP} {\bfseries 01} (1999) 003} [\href{https://arxiv.org/abs/hep-th/9807137}{{\ttfamily hep-th/9807137}}].

\bibitem{Taylor:2017dly}
M.~Taylor, \emph{{Generalized conformal structure, dilaton gravity and SYK}}, \href{https://doi.org/10.1007/JHEP01(2018)010}{\emph{JHEP} {\bfseries 01} (2018) 010} [\href{https://arxiv.org/abs/1706.07812}{{\ttfamily 1706.07812}}].

\bibitem{Plebanski:1976gy}
J.~F. Plebanski and M.~Demianski, \emph{{Rotating, charged, and uniformly accelerating mass in general relativity}}, \href{https://doi.org/10.1016/0003-4916(76)90240-2}{\emph{Annals Phys.} {\bfseries 98} (1976) 98}.

\bibitem{Griffiths:2006tk}
J.~B. Griffiths, P.~Krtous and J.~Podolsky, \emph{{Interpreting the C-metric}}, \href{https://doi.org/10.1088/0264-9381/23/23/008}{\emph{Class. Quant. Grav.} {\bfseries 23} (2006) 6745} [\href{https://arxiv.org/abs/gr-qc/0609056}{{\ttfamily gr-qc/0609056}}].

\bibitem{Brown:1986nw}
J.~D. Brown and M.~Henneaux, \emph{{Central Charges in the Canonical Realization of Asymptotic Symmetries: An Example from Three-Dimensional Gravity}}, \href{https://doi.org/10.1007/BF01211590}{\emph{Commun. Math. Phys.} {\bfseries 104} (1986) 207}.

\bibitem{Birmingham:1998nr}
D.~Birmingham, \emph{{Topological black holes in Anti-de Sitter space}}, \href{https://doi.org/10.1088/0264-9381/16/4/009}{\emph{Class. Quant. Grav.} {\bfseries 16} (1999) 1197} [\href{https://arxiv.org/abs/hep-th/9808032}{{\ttfamily hep-th/9808032}}].

\bibitem{Mann:1997iz}
R.~B. Mann, \emph{{Topological black holes: Outside looking in}}, {\emph{Annals Israel Phys. Soc.} {\bfseries 13} (1997) 311} [\href{https://arxiv.org/abs/gr-qc/9709039}{{\ttfamily gr-qc/9709039}}].

\bibitem{Banados:1997df}
M.~Banados, \emph{{Constant curvature black holes}}, \href{https://doi.org/10.1103/PhysRevD.57.1068}{\emph{Phys. Rev. D} {\bfseries 57} (1998) 1068} [\href{https://arxiv.org/abs/gr-qc/9703040}{{\ttfamily gr-qc/9703040}}].

\end{thebibliography}\endgroup

\end{document}